\documentclass[11pt]{amsart}
\usepackage{amsbsy,amsfonts}
\usepackage{amsaddr}
\usepackage{latexsym,eucal,amsmath,amsthm}
\usepackage{amssymb}
\usepackage{graphicx}
\usepackage{array}
\usepackage[colorlinks=true,citecolor=blue,linkcolor=black]{hyperref}
\usepackage{tikz}
\usepackage{environ}
\usepackage{etoolbox}
\usepackage{graphicx}
\usepackage{adjustbox}
\usepackage{appendix}
\usepackage{color}

\AtBeginEnvironment{subappendices}{
    
                            }
\AtBeginEnvironment{subappendices}{
    
                            }
\newlength{\myl}
\let\origequation=\equation
\let\origendequation=\endequation
\let\equation\relax
\let\endequation\relax
\NewEnviron{equation}{
  \settowidth{\myl}{$\BODY$}                         \origequation
  \ifdimcomp{\the\linewidth}{>}{\the\myl}
  {\ensuremath{\BODY}}                               {\resizebox{\linewidth}{!}{\ensuremath{\BODY}}}    \origendequation
}

\newtheorem*{hyp}{Hypothesis}

\theoremstyle{definition}

\theoremstyle{remark}
\newtheorem{remark}{Remark}
\theoremstyle{proposition}

\theoremstyle{lemma}

\theoremstyle{corollary}

\numberwithin{equation}{section}
\numberwithin{lemma}{section}
\numberwithin{remark}{section}

\begin{document}
\title[Bound States in $n$th-order Schr\"{o}dinger equations]{%
Quantization Condition of the Bound States in $n$th-Order Schr\"{o}dinger Equations}
\author{Xiong Fan}
%\address{Hong Kong University of Science and Technology, Hong Kong, China}
\email{xfanam@mail.ust.hk}
\address{Hong Kong University of Science and Technology, Hong Kong, China}
% \address{National University of Defense Technology, Changsha, China}
\begin{abstract}
We prove a general approximate quantization rule $
\int_{L_{E}}^{R_{E}}k_0(x)$ $dx=(N+\frac{1}{2})\pi $ or $
\oint k_0(x)$ $dx=(2N+1)\pi $ (including both forward and backward processes) for the bound states in
the potential well of the $n$th-order Schr\"{o}dinger equations $ e^{-i\pi n/2}{{}\frac{d^n\Psi(x)}{d x^n} }
=[E-{} V(x)]\Psi(x) ,$ where ${} k_0(x)=(E-V(x) )^{1/n}$ with $N\in\mathbb{N}_{0} $, $n$ is an even natural number, and $L_{E}$ and $R_{E}$ the boundary
points between the classically forbidden regions and the allowed region. The
only hypothesis is that all exponentially growing components are negligible,
which is appropriate for not narrow wells. Applications including
the Schr\"{o}dinger equation and Bogoliubov-de Gennes equation will be
discussed.
\end{abstract}

\date{}
\maketitle

\section{Introduction}

The WKB approximation is a classic in quantum mechanics \cite%
{weinberg2015lectures,griffiths2018}. It is a powerful technique for
obtaining approximate wave functions and bound state energies in the Schr%
\"{o}dinger equation. The foundation is that potentials change much more
slowly than wave functions.
% The traditional way of the WKB approximation is
%expanding the wave function of the Schr\"{o}dinger equation in powers of $%
%\hbar$, and when $\hbar\rightarrow0$, the result reduces to classical
%physics. For this reason, the WKB approximation is also named as a
%semiclassical method.
The WKB method adds an additional $\frac{1}{2}h$ to the Bohr-Sommerfeld quantization rule, leading to $\oint pdq=(N+\frac{1}{2})h$, which exactly solves the quantum harmonic oscillator; in addition, the exact quantization condition is recently derived by Ma and Xu \cite{Ma_2005}. The WKB method is conventionally attributed to Wenzel
\cite{wentzel1926}, Kramer \cite{kramers1926}, and Brillouin \cite{brillouin1926}, who founded the theory to understand the Schr\"{o}dinger equation, and Jeffreys' work \cite{jeffreys1925} completed before the Schr\"{o}dinger equation also contributes to this method. One essence of their
works is the treatment to turning points, which will also be exploited in
this paper. Recent developments in WKB include supersymmetry WKB (SWKB) \cite{comtet1985,khare1985,dutt1991,cooper1995} and semiclassical quantum gravity \cite{banks1985,singh1989}. SWKB gives exact eigenvalues with known ground-state wave functions for shape invariant potentials. Semiclassical quantum
gravity utilizes WKB to solve the famous Wheeler-DeWitt equation. The WKB
approximation is commonly applied to second-order differential equations,
and it is noticed that this method can be used in higher-order differential
equations \cite{bender1999}. Nevertheless, the WKB approximation in
higher-order differential equations did not advance.

\begin{figure}[htb]
\centering
\includegraphics[width=9cm]{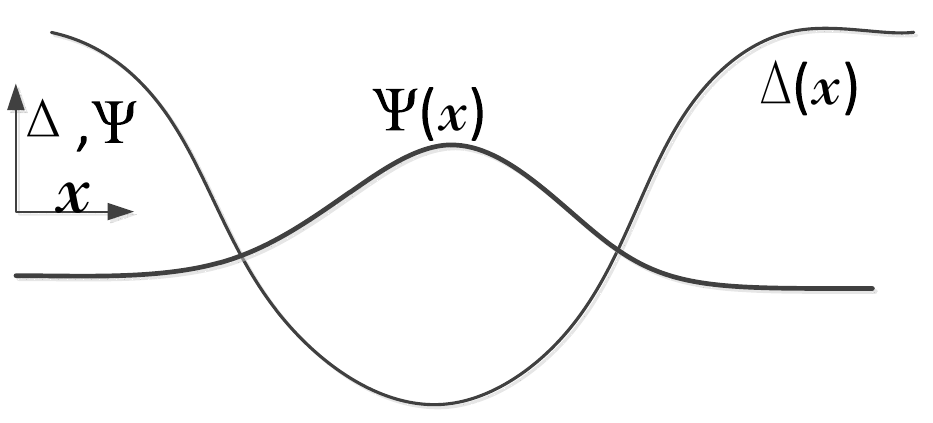}
\caption{Illustration of a bound state $\Psi(x)$ in a potential well ${{} V}(x)$. The bound state wave function $\Psi(x)$ dissipates gradually within the barrier ${{} V(x) }$.}\label{demo}
\end{figure}

{{} In Ref. \cite{fan2022}, we solved a fourth-order ordinary differential equation for a Bogoliubov-de Gennes equation with parabolic dispersions, and observed that the quantizaton condition of the bound states coincides with that of the Schr\"{o}dinger equation.} In this paper, we will discuss the bound states in general higher-order differential
equations, and prove the general quantization condition shown in the
abstract. An illustration of the bound state in the potential well is drawn in Fig. \ref{demo}. It is found that the general quantization condition is invariant except for the wave vector which changes according to the order of differential equations. Thus, we generalize the WKB result to higher-order differential equations, and the Schr\"{o}dinger equations are merely a class of important examples in our proof. We shall see that both phase
factors and Bessel functions, which are significant for theoretical physics,
play pivotal roles in the proof.

Before the proof, let us see the brief procedure first. (1) We find the
asymptotic behavior of the wave functions; (2) Use the Frobenius method to
find the wave function near turning points, and get $\Psi _{j}(x)$; (3)
Continue with the wave function obtained in the second step, {{} change the
variable} $x$ of the wave function to an integral $\phi _{0}$, i.e., $\Psi _{j}(x)\rightarrow \Psi _{j}(\phi _{0})$; (4) Use the Bessel functions to represent $\Psi _{j}(\phi _{0})$, and use $\Psi _{\kappa
}(\phi _{0})$ to represent the asymptotic behavior of $\Psi _{j}(\phi
_{0})$; (5) Use the continuous condition at the turning points to find the
relationship between coefficients in $\Psi =\sum_{j}\alpha _{j}\Psi _{j}$ on two sides of a specific turning point; (6) Let all exponentially growing
components vanish, and prove another relationship between coefficients; (7)
Finally, compare with the asymptotic behavior acquired in the first step,
and then obtain the quantization condition. After the proof, we will give
typical physical applications. Last but not least, in the remarks, we
will distinguish between odd $n$ and even $n$ cases, and discuss some crucial
calculation details, conjectures, etc. Symbols with similar appearances
are collected in Appendix \ref{{symbols}}.

\section{\protect\bigskip Bound states in $e^{-i\protect\pi n/2}\protect%
{{}\frac{d^n\Psi(x)}{d x^n} } =(E-{{} V(x) } )\Psi $}
{{} The methods of solving the case with $n=4,8,12,\cdots$ and the case with $n=2,6,10,\cdots $ are different. We discuss the former in Sec. 2.1 and the latter in Sec. 2.2. The value $e^{-i\protect\pi n/2}$ is different for two cases such that the free particle states $e^{\pm ikx}$ survive inside the potential well where $E>V(x)$.}
\subsection{$n=2\ast (2n_{c})$}

$n_{c}$ is a positive integer. The equation becomes
\begin{equation}
{{}\frac{d^n\Psi(x)}{d x^n} } =[E-{} V (x)]\Psi ,  \label{psifun}
\end{equation}
where $E$ is a constant, and ${{} V(x) } $ changes smoothly with $x.$ Now we
assume that ${} V (x)$ has a well shape. $E-{} V (x)>0$ in the
classically allowed region $L_{E}<x<R_{E}$, and $E-{} V (x)<0$ in the
classically forbidden regions $x<L_{E}$ and $x>R_{E}$. $L_{E}$ and $R_{E}$
are the left and right boundary points between the classically allowed and
forbidden regions. Following Wenzel. \emph{et al.} and Weinberg's perspective and another
paper \cite{fan2022}, we start from the asymptotic behavior to
unearth the bound states of this equation.

\subsubsection{ Asymptotic behavior\label{asymptoticb}}

When $x>R_{E},$ define $\kappa _{0}=[{} V (x)-E]^{1/n}$ with arg$(\kappa
_{0})=0,$ the equation $ e^{-i\pi n/2}{{}\frac{d^n\Psi(x)}{d x^n} }
=[E-{} V(x)]\Psi(x) $ becomes ${{}\frac{d^n\Psi(x)}{d x^n} } +\kappa _{0}^{n}\Psi
=0.$ If ${} V (x)$ {{} were} a constant, the wave function $\Psi $ has the
solution in the form $e^{\kappa x},$ and $\kappa $ has $n$ possible values: $%
e^{\pm i\frac{\pi }{n}\ast 1}\kappa _{0},e^{\pm i\frac{\pi }{n}\ast 3}\kappa
_{0},\cdots ,e^{\pm i\frac{\pi }{n}\ast (n-1)}\kappa _{0}.$ {{} Now we assume
that ${} V (x)$ changes slowly, and the wave function takes the hypothesized form $\Psi ={} f (x)\psi(x) $ with $\psi=e^{\int \kappa dx}$. $f(x)$ is a slowly varying amplitude, and $\psi(x) $ describes the exponetial behavior of the wave function.} Substituting the hypothesized form into Eq. (\ref%
{psifun}) and only keeping first-order derivative terms, we have
\begin{equation}
{} n(\frac{df}{dx})\kappa +\frac{n(n-1)}{2}(\frac{d\kappa}{dx} )f=0.  \label{kappa0}
\end{equation}%
{{} In Appendix \ref{firsta}, we give conditions under which higher-order terms can be neglected. Since we need to deal with the $(n-1)$th-order derivative of $\kappa$ in Appendix \ref{firsta}, the potential function $V(x)$ is required to be a $C^{n-1}-$function.} Therefore, $f\propto \kappa ^{\frac{1-n}{2}}.$ Eq. (\ref{kappa0}) is valid
only if other terms are negligible; see Ref. \cite{SM}. To be
concise, we omit similar discussions for other regions. {{} In this part, we have obtained asymptotic behaviors of the wave function far away from the turning points.} We will return to
approach this asymptotic formula from the wave functions obtained near the
turning points.

\subsubsection{Turning points}

Near the point $R_{E}$, $f\rightarrow \infty$ {{} since $\kappa\rightarrow0$}. Consequently,
the hypothesized form is not appropriate. To tackle this range
separately, use the Taylor series to expand ${} V (x),$ and make the linear
approximation: ${} V (x)=E+{} V
_{x=R_{E}}^{(1)}(x-R_{E})+O[(x-R_{E})^{2}]$ with {{} $V
_{x=R_{E}}^{(1)}$ the first-order derivative of $V (x)$ at $x=R_{E}$}. Near $R_{E}$, we have
\begin{eqnarray}
{{}\frac{d^n\Psi(x)}{d x^n} } &=&[-{} V _{x=R_{E}}^{(1)}(x-R_{E})]\Psi ,  \notag
\\
\text{or }{{}\frac{d^n\Psi(x^{\prime })}{d x^{\prime n}} } &=&[-{} V _{x^{\prime
}=0}^{(1)}x^{\prime }]\Psi ,  \label{substitution}
\end{eqnarray}
if we do the substitution $x^{\prime }=x-R_{E}.$ For simplicity, we still
use $x$ to denote $x^{\prime }.$ We use the Frobenius method to solve this
equation. {{} Since the coefficient $[-{} V _{x^{
}=0}^{(1)}x^{ }]$ is analytic in the whole domain, we assume $\Psi =\sum\limits_{m=0}^{\infty }a_{m}x^{m}$ \cite{ince1956}}. We get the
recursive relationship
\begin{equation}
\frac{a_{m+n+1}}{a_{m}}=\frac{-{} V _{x=R_{E}}^{(1)}}{(m+n+1)(m+n+1-1)%
\cdots \lbrack m+n+1-(n-1)]}.  \label{ite1}
\end{equation}
Based on the recursive relationship, $\Psi $ has $n$ independent solutions:
\begin{eqnarray}
\Psi _{0} &:&a_{0}\neq 0,a_{1}=a_{2}=\cdots =a_{n-1}=0;  \label{psij} \\
\Psi _{1} &:&a_{1}\neq 0,a_{0}=a_{2}=\cdots =a_{n-1}=0;  \notag \\
&&\vdots  \notag \\
\Psi _{j} &:&a_{j}\neq 0,a_{0}=a_{1}=\cdots =a_{n-1}=0;  \notag \\
&&\vdots  \notag \\
\Psi _{n-1} &:&a_{n-1}\neq 0,a_{0}=a_{1}=\cdots =a_{n-2}=0.  \notag
\end{eqnarray}
In short,
\begin{equation}
{} \Psi _{j}=\sum\limits_{m_{}=j}^{{\infty}}a_{m}x^{m} \text{\,\,with mod}(m-j,n+1)=0.
\end{equation}

{{} In this part, we have expressed the $n$ independent wave functions near the turning point as power series.}

\subsubsection{$\Psi (x)\rightarrow \Psi (\protect\phi _{0})$}

{{} Now we use $\phi_{0}$ as the variable of the wave function, and do some
approximations to obtain Eq. (\ref{psijaa}). Define $\phi
_{0}=\int_{0}^{x}\kappa _{0}dx^{\prime }=\frac{n}{n+1}[{} V
_{x=R_{E}}^{(1)}]^{1/n}x^{\frac{n+1}{n}}.$ In Secs. \ref{bessel} and \ref{quancon}, we see that $\phi
_{0}$ corresponds to a phase inside the potential well.} We transform $\Psi _{0}$
first.
\begin{eqnarray}
\Psi _{0} &=&\sum\limits_{m=k(n+1),k=0,1,2,\cdots
}^{{}}a_{m}x^{m}=\sum\limits_{k=0,1,2,\cdots }^{{}}a_{k}x^{k(n+1)} \\
\text{with}\quad\frac{a_{k}}{a_{k-1}} &=&\frac{-{} V _{x=R_{E}}^{(1)}}{%
[k(n+1)][k(n+1)-1]\cdots \lbrack k(n+1)-(n-1)]}.  \notag
\end{eqnarray}
Therefore, %\newline
$a_{k}=\frac{[-{} V _{x=R_{E}}^{(1)}]^{k}a_{0}}{{\prod_{k^{\prime
}=1,2,\cdots ,k}[k^{\prime }(n+1)][k^{\prime }(n+1)-1]\cdots \lbrack
k^{\prime }(n+1)-(n-1)]}} $. Since $\phi _{0}^{n}=(\frac{n}{n+1}%
)^{n}[{{} V} _{x=R_{E}}^{(1)}]x^{n+1},$ replacing $x$ by $\phi _{0},$ we
have
\begin{equation}
\Psi _{0}=\sum\limits_{k=0,1,2,\cdots
}^{{}}a_{k}x^{k(n+1)}=\sum\limits_{k=0,1,2,\cdots }^{{}}a_{k}\phi _{0}^{kn}(%
\frac{n}{n+1})^{-kn}[{} V _{x=R_{E}}^{(1)}]^{-k}.  \label{kp1}
\end{equation}%
Let $\overline{a}_{k}=a_{k}(\frac{n}{n+1})^{-kn}[{{} V}
_{x=R_{E}}^{(1)}]^{-k}=\frac{(-1)^{k}a_{0}}{\prod_{k^{\prime }=1,2,\cdots
,k}[k^{\prime }n][k^{\prime }n-1\frac{n}{n+1}]\cdots \lbrack k^{\prime
}n-(n-1)\frac{n}{n+1}]}. $ Change the notation $\overline{a}%
_{k}\rightarrow \overline{a}_{kn}.$ In short, $\Psi
_{0}=\sum\limits_{k=0,1,2,\cdots }^{{}}\overline{a}_{kn,0}^{{}}\phi
_{0}^{kn}.$

Similarly, $\Psi _{j}=\sum\limits_{k=0,1,2,\cdots }^{{}}\overline{a}%
_{kn,j}\phi _{0}^{kn+j\frac{n}{n+1}}$ with
\begin{equation}
\overline{a}_{kn,j}=\frac{(-1)^{k}a_{0,j}}{\prod_{k^{\prime }=1,2,\cdots
,k}[k^{\prime }n+j\frac{n}{n+1}][k^{\prime }n+j\frac{n}{n+1}-1\frac{n}{n+1}%
]\cdots \lbrack k^{\prime }n+j\frac{n}{n+1}-(n-1)\frac{n}{n+1}]}.
\end{equation}
In the following, we do approximations to $\overline{a}_{kn,j},$ then $\Psi $
can be represented by Bessel functions. We take $\Psi _{0}$ as an example. Change the form of $\overline{a}_{kn,0}$ first:
\begin{eqnarray}
\overline{a}_{kn,0} &=&\frac{(-1)^{k}a_{0,0}}{\prod_{k^{\prime }=1,2,\cdots
,k}[k^{\prime }n][k^{\prime }n-1\frac{n}{n+1}]\cdots \lbrack k^{\prime
}n-(n-1)\frac{n}{n+1}]}  \notag \\
%&=&\frac{(-1)^{k}a_{0,0}}{%
%\begin{array}{c}
%{\prod_{k^{\prime }=1,2,\cdots ,k}2^{n}[k^{\prime }n/2][k^{\prime }n/2-1%
%\frac{n/2}{n+1}][k^{\prime }n/2-2\frac{n/2}{n+1}]\cdots \lbrack k^{\prime
%}n/2-(n-3)\frac{n/2}{n+1}]\ast } \\
%{[k^{\prime }n/2-(n-2)\frac{n/2}{n+1}][k^{\prime }n/2-(n-1)\frac{n/2}{n+1}]}%
%\end{array}%
%} \\
&=&\frac{(-1)^{k}a_{0,0}}{{%
\begin{array}{c}
\prod_{k^{\prime }=1,2,\cdots ,k}2^{n}[k^{\prime }n/2][k^{\prime }n/2-\frac{1%
}{n+1}-\frac{n/2-1}{n+1}][k^{\prime }n/2-1+\frac{1}{n+1}]\cdots \ast \\
\lbrack k^{\prime }n/2-(n/2-2)-\frac{2}{n+1}][k^{\prime }n/2-(n/2-1)+\frac{%
n/2-1}{n+1}][k^{\prime }n/2-(n/2-1)-\frac{1}{n+1}]%
\end{array}%
}}.  \notag
\end{eqnarray}
Let us see the odd terms $[k^{\prime }n/2-(n-r)\frac{n/2}{n+1}]$ in the
denominator for a fixed $k^{\prime }$ with $r=2,4,..., n$:
\begin{equation}
\left\{
\begin{array}{c}
\lbrack k^{\prime }n/2],[k^{\prime }n/2-1+\frac{1}{n+1}],\cdots ,[k^{\prime
}n/2-(n-r)/2+\frac{(n-r)/2}{n+1}],\cdots , \\
\lbrack k^{\prime }n/2-(n-4)/2+\frac{(n-4)/2}{n+1}],[k^{\prime }n/2-(n-2)/2+%
\frac{(n-2)/2}{n+1}]%
\end{array}%
\right\} ;
\end{equation}
and the even terms $[k^{\prime }n/2-(n-t)\frac{n/2}{n+1}]$  with $t=1,3,..., n-1$:
\begin{equation}
\left\{
\begin{array}{c}
\lbrack k^{\prime }n/2-\frac{1}{n+1}-\frac{(n-2)/2}{n+1}],[k^{\prime }n/2-1-%
\frac{(n-4)/2}{n+1}],\cdots ,[k^{\prime }n/2-(n-t-1)/2-\frac{1}{n+1}-\frac{%
(t-1)/2}{n+1}], \\
\cdots ,[k^{\prime }n/2-(n-4)/2-\frac{1}{n+1}-\frac{1}{n+1}],[k^{\prime
}n/2-(n-2)/2-\frac{1}{n+1}]%
\end{array}%
\right\} .
\end{equation}
When $\frac{(n-r)/2}{n+1}=\frac{(t-1)/2}{n+1},$ i.e. $r+t=n+1,$ two terms
form a pair, and we do the following approximation for such a pair:
\begin{align}
\lbrack k^{\prime }n/2-(n-r)/2+\frac{(n-r)/2}{n+1}]& [k^{\prime
}n/2-(n-t-1)/2-\frac{1}{n+1}-\frac{(t-1)/2}{n+1}]\approx \\
& [k^{\prime }n/2-(n-r)/2][k^{\prime }n/2-(n-t-1)/2-\frac{1}{n+1}].  \notag
\end{align}
The approximation is redundant for the $n=2$ case.
Each term is paired with another term in the symmetric position,
after doing the approximation to all pairs for each $k^{\prime }$, we have
\begin{equation}
\overline{a}_{kn,0}\approx \frac{(-1)^{k}}{2^{kn}(kn/2)!\Gamma (kn/2-\frac{1%
}{n+1}+1)}
\end{equation}%
with $a_{0,0}=\frac{1}{\Gamma (1)\Gamma (0-\frac{1}{n+1}+1)}. $
{{}  Therefore, $\Psi _{0}=\sum\limits_{k=0,1,2,\cdots }^{{}}\overline{a}%
_{kn,0}\phi _{0}^{kn}$, and we always adopt the approximated value of $\overline{a}_{kn,j}$ in the following.}

Before doing similar approximations to $\overline{a}_{kn,j},$ we define $%
2b=\left\{
\begin{array}{ll}
j, & \hbox{mod$(j,2)=0;$} \\
j-1, & \hbox{mod$(j-1,2)=0.$}%
\end{array}%
\right. $ I.e., $j=0,1$ correspond to $b=0;$ $j=2,3$ correspond to $b=1,$
and so forth. In other words, $b$ is like a staircase, adding one step every
time $j$ moves $2$ units starting from the origin. Besides, defining $\nu
_{j}=j\frac{n}{n+1}-2b_{j}-\frac{1}{n+1},$ we have $\nu _{j}=\left\{
\begin{array}{ll}
-\frac{j+1}{n+1}, & \hbox{\text{ mod}(j,2)=0;} \\
\frac{n-j}{n+1}, & \hbox{\text{ mod}(j,2)=1,}%
\end{array}%
\right. $ which will be the subscript parameter of the Bessel function $%
I_{\nu _{j}}.$

We discuss even $j$ and odd $j$ separately. When $j$ is even,
\begin{eqnarray}
\overline{a}_{kn,j} &=&\frac{(-1)^{k}a_{0,j}}{%
\begin{array}{c}
\prod_{k^{\prime }=1,2,\cdots ,k}[k^{\prime }n+j\frac{n}{n+1}][k^{\prime }n+j%
\frac{n}{n+1}-1\frac{n}{n+1}][k^{\prime }n+j\frac{n}{n+1}-2\frac{n}{n+1}%
]\cdots \ast \\
\lbrack k^{\prime }n+j\frac{n}{n+1}-(n-2)\frac{n}{n+1}][k^{\prime }n+j\frac{n%
}{n+1}-(n-1)\frac{n}{n+1}]%
\end{array}%
}  \notag \\
&=&\frac{(-1)^{k}a_{0,j}}{%
\begin{array}{c}
\prod_{k^{\prime }=1,2,\cdots ,k}2^{n}[k^{\prime }n/2+j\frac{n/2}{n+1}%
][k^{\prime }n/2+j\frac{n/2}{n+1}-1\frac{n/2}{n+1}][k^{\prime }n/2+j\frac{n/2%
}{n+1}-2\frac{n/2}{n+1}]\cdots \ast \\
\lbrack k^{\prime }n/2+j\frac{n/2}{n+1}-(n-2)\frac{n/2}{n+1}][k^{\prime
}n/2+j\frac{n/2}{n+1}-(n-1)\frac{n/2}{n+1}]%
\end{array}%
}.
\end{eqnarray}%
For the odd terms in the denominator for a fixed $k^{\prime }:\left( r\text{
is even}\right) $
\begin{eqnarray}
&&\left\{
\begin{array}{c}
\lbrack k^{\prime }n/2+j\frac{n/2}{n+1}],[k^{\prime }n/2+j\frac{n/2}{n+1}-2%
\frac{n/2}{n+1}],\cdots ,[k^{\prime }n/2+j\frac{n/2}{n+1}-(n-r)\frac{n/2}{n+1%
}], \\
\cdots, \lbrack k^{\prime }n/2+j\frac{n/2}{n+1}-(n-4)\frac{n/2}{n+1}%
],[k^{\prime }n/2+j\frac{n/2}{n+1}-(n-2)\frac{n/2}{n+1}]%
\end{array}%
\right\} \\
%&=&\left\{
%\begin{array}{c}
%\lbrack k^{\prime }n/2+j\frac{n/2}{n+1}],[k^{\prime }n/2+(j-2)\frac{n/2}{n+1}%
%],\cdots ,\{k^{\prime }n/2+[j-(n-r)]\frac{n/2}{n+1}\}, \\
%\cdots, \{k^{\prime }n/2+[j-(n-4)]\frac{n/2}{n+1}\},\{k^{\prime
%}n/2+[j-(n-2)]\frac{n/2}{n+1}\}%
%\end{array}%
%\right\}  \notag \\
%&=&\left\{
%\begin{array}{c}
%\lbrack k^{\prime }n/2+\frac{j}{2}(1-\frac{1}{n+1})],[k^{\prime }n/2+\frac{%
%(j-2)}{2}(1-\frac{1}{n+1})],\cdots ,\{k^{\prime }n/2+\frac{[j-(n-r)]}{2}(1-%
%\frac{1}{n+1})\}, \\
%\cdots, \{k^{\prime }n/2+\frac{[j-(n-4)]}{2}(1-\frac{1}{n+1})\},\{k^{\prime
%}n/2+\frac{[j-(n-2)]}{2}(1-\frac{1}{n+1})\}%
%\end{array}%
%\right\}  \notag \\
&=&\left\{
\begin{array}{c}
\lbrack k^{\prime }n/2+\frac{j}{2}-\frac{j}{2}\frac{1}{n+1}],[k^{\prime
}n/2+(\frac{j}{2}-1)-\frac{(j-2)}{2}\frac{1}{n+1})], \\
\cdots ,\{k^{\prime }n/2+[\frac{j}{2}-\frac{(n-r)}{2}]-\frac{[j-(n-r)]}{2}%
\frac{1}{n+1}\},\cdots , \\
\{k^{\prime }n/2+[\frac{j}{2}-\frac{(n-4)}{2}]-\frac{[j-(n-4)]}{2}\frac{1}{%
n+1})\},\{k^{\prime }n/2+[\frac{j}{2}-\frac{(n-2)}{2}]-\frac{[j-(n-2)]}{2}%
\frac{1}{n+1}\}%
\end{array}%
\right\}.  \notag
\end{eqnarray}%
For the even terms in the denominator for a fixed $k^{\prime }:\left( t\text{
is odd}\right) $
\begin{eqnarray}
&&\left\{
\begin{array}{c}
\lbrack k^{\prime }n/2+j\frac{n/2}{n+1}-\frac{n/2}{n+1}],[k^{\prime }n/2+j%
\frac{n/2}{n+1}-3\frac{n/2}{n+1}],\cdots ,[k^{\prime }n/2+j\frac{n/2}{n+1}%
-(n-t)\frac{n/2}{n+1}], \\
\cdots ,[k^{\prime }n/2+j\frac{n/2}{n+1}-(n-3)\frac{n/2}{n+1}],[k^{\prime
}n/2+j\frac{n/2}{n+1}-(n-1)\frac{n/2}{n+1}]%
\end{array}%
\right\} \\
%&=&\left\{
%\begin{array}{c}
%\lbrack k^{\prime }n/2+(j-1)\frac{n/2}{n+1}],[k^{\prime }n/2+(j-3)\frac{n/2}{%
%n+1}],\cdots ,\{k^{\prime }n/2+[j-(n-t)]\frac{n/2}{n+1}\}, \\
%\cdots ,\{k^{\prime }n/2+[j-(n-3)]\frac{n/2}{n+1}\},\{k^{\prime
%}n/2+[j-(n-1)]\frac{n/2}{n+1}\}%
%\end{array}%
%\right\}  \notag \\
&=&\left\{
\begin{array}{c}
\lbrack k^{\prime }n/2+\frac{j}{2}-\frac{j+1}{n+1}+\frac{j-n+2}{2(n+1)}%
],[k^{\prime }n/2+(\frac{j}{2}-1)-\frac{j+1}{n+1}+\frac{j-n+4}{2(n+1)}], \\
\cdots ,\{k^{\prime }n/2+(\frac{j}{2}-\frac{n-t-1}{2})-\frac{j+1}{n+1}+\frac{%
j-t+1}{2(n+1)}\},\cdots , \\
\{k^{\prime }n/2+(\frac{j}{2}-\frac{n-4}{2})-\frac{j+1}{n+1}+\frac{j-2}{%
2(n+1)}\},\{k^{\prime }n/2+(\frac{j}{2}-\frac{n-2}{2})-\frac{j+1}{n+1}+\frac{%
j}{2(n+1)}\}%
\end{array}%
\right\} .  \notag
\end{eqnarray}%
Two terms form a pair if $\frac{[j-(n-r)]}{2}\frac{1}{n+1}=\frac{j-t+1}{%
2(n+1)},$ which is $r+t=n+1.$ For each pair, we do the approximation:
\begin{align}
\{k^{\prime }n/2+[\frac{j}{2}-\frac{(n-r)}{2}]-&\frac{[j-(n-r)]}{2}\frac{1}{%
n+1}\} \{k^{\prime }n/2+(\frac{j}{2}-\frac{n-t-1}{2})-\frac{j+1}{n+1}+\frac{%
j-t+1}{2(n+1)}\}\approx  \notag \\
& \{k^{\prime }n/2+[\frac{j}{2}-\frac{(n-r)}{2}]\}\{k^{\prime }n/2+(\frac{j}{%
2}-\frac{n-t-1}{2})-\frac{j+1}{n+1}\}.
\end{align}%
After doing the approximation to all pairs in each $k^{\prime },$ we have
\begin{eqnarray}
\overline{a}_{kn,j} &\approx &\frac{(-1)^{k}}{2^{kn}(kn/2+\frac{j}{2}%
)!\Gamma (kn/2+\frac{j}{2}-\frac{j+1}{n+1}+1)}
=\frac{(-1)^{k}}{2^{kn}(kn/2+b_{j})!\Gamma (kn/2+b_{j}-\frac{j+1}{n+1}+1)}
\notag
\end{eqnarray}%
with $a_{0,j}=\frac{1}{\Gamma (\frac{j}{2}+1)\Gamma (\frac{j}{2}-\frac{j+1}{%
n+1}+1)}. $ Now we have $\Psi _{j}=\sum\limits_{k=0,1,2,\cdots }^{{}}%
\overline{a}_{kn,j}\phi _{0}^{kn+j\frac{n}{n+1}}$ with the approximated $\overline{a}_{kn,j}$.
%=\sum\limits_{k=0,1,2,\cdots
%}^{{}}\overline{a}_{kn,j}\phi _{0}^{kn+j-\frac{j}{n+1}}.

When $j$ is odd,
\begin{eqnarray}
\overline{a}_{kn,j} &=&\frac{(-1)^{k}a_{0,j}}{%
\begin{array}{c}
\prod_{k^{\prime }=1,2,\cdots ,k}[k^{\prime }n+j\frac{n}{n+1}][k^{\prime }n+j%
\frac{n}{n+1}-1\frac{n}{n+1}][k^{\prime }n+j\frac{n}{n+1}-2\frac{n}{n+1}%
]\cdots \ast \\
\lbrack k^{\prime }n+j\frac{n}{n+1}-(n-2)\frac{n}{n+1}][k^{\prime }n+j\frac{n%
}{n+1}-(n-1)\frac{n}{n+1}]%
\end{array}%
}  \notag \\
&=&\frac{(-1)^{k}a_{0,j}}{%
\begin{array}{c}
\prod_{k^{\prime }=1,2,\cdots ,k}2^{n}[k^{\prime }n/2+j\frac{n/2}{n+1}%
][k^{\prime }n/2+j\frac{n/2}{n+1}-1\frac{n/2}{n+1}][k^{\prime }n/2+j\frac{n/2%
}{n+1}-2\frac{n/2}{n+1}]\cdots \ast \\
\lbrack k^{\prime }n/2+j\frac{n/2}{n+1}-(n-2)\frac{n/2}{n+1}][k^{\prime
}n/2+j\frac{n/2}{n+1}-(n-1)\frac{n/2}{n+1}]%
\end{array}%
}.
\end{eqnarray}
For the even terms in the denominator for a fixed $k^{\prime }:\left( t\text{
is odd}\right) $
\begin{eqnarray}
&&\left\{
\begin{array}{c}
\lbrack k^{\prime }n/2+j\frac{n/2}{n+1}-\frac{n/2}{n+1}],[k^{\prime }n/2+j%
\frac{n/2}{n+1}-3\frac{n/2}{n+1}],\cdots ,[k^{\prime }n/2+j\frac{n/2}{n+1}%
-(n-t)\frac{n/2}{n+1}], \\
\cdots ,[k^{\prime }n/2+j\frac{n/2}{n+1}-(n-3)\frac{n/2}{n+1}],[k^{\prime
}n/2+j\frac{n/2}{n+1}-(n-1)\frac{n/2}{n+1}]%
\end{array}%
\right\} \\
%&=&\left\{
%\begin{array}{c}
%\lbrack k^{\prime }n/2+(j-1)\frac{n/2}{n+1}],[k^{\prime }n/2+(j-3)\frac{n/2}{%
%n+1}],\cdots ,\{k^{\prime }n/2+[j-(n-t)]\frac{n/2}{n+1}\}, \\
%\cdots ,\{k^{\prime }n/2+[j-(n-3)]\frac{n/2}{n+1}\},\{k^{\prime
%}n/2+[j-(n-1)]\frac{n/2}{n+1}\}%
%\end{array}%
%\right\}  \notag \\
&=&\left\{
\begin{array}{c}
\lbrack k^{\prime }n/2+\frac{j-1}{2}-\frac{j-1}{2(n+1)}],[k^{\prime }n/2+%
\frac{j-3}{2}-\frac{j-3}{2(n+1)}],\cdots ,\{k^{\prime }n/2+\frac{j-(n-t)}{2}-%
\frac{j-(n-t)}{2(n+1)}\}, \\
\cdots ,\{k^{\prime }n/2+\frac{j-(n-3)}{2}-\frac{j-(n-3)}{2(n+1)}%
\},\{k^{\prime }n/2+\frac{j-(n-1)}{2}-\frac{j-(n-1)}{2(n+1)}\}%
\end{array}%
\right\} .  \notag
\end{eqnarray}%
For the odd terms in the denominator for a fixed $k^{\prime }:\left( r\text{
is even}\right) $
\begin{eqnarray}
&&\left\{
\begin{array}{c}
\lbrack k^{\prime }n/2+j\frac{n/2}{n+1}],[k^{\prime }n/2+j\frac{n/2}{n+1}-2%
\frac{n/2}{n+1}],\cdots ,[k^{\prime }n/2+j\frac{n/2}{n+1}-(n-r)\frac{n/2}{n+1%
}], \\
\cdots , \lbrack k^{\prime }n/2+j\frac{n/2}{n+1}-(n-4)\frac{n/2}{n+1}%
],[k^{\prime }n/2+j\frac{n/2}{n+1}-(n-2)\frac{n/2}{n+1}]%
\end{array}%
\right\} \\
&=&\left\{
\begin{array}{c}
\lbrack k^{\prime }n/2+\frac{j-1}{2}+\frac{n-j}{n+1}+\frac{j-(n-1)}{2(n+1)}%
],[k^{\prime }n/2+\frac{j-3}{2}+\frac{n-j}{n+1}+\frac{j-(n-3)}{2(n+1)}], \\
\cdots ,[k^{\prime }n/2+\frac{j-[(n-r)+1]}{2}+\frac{n-j}{n+1}+\frac{j-(r-1)}{%
2(n+1)}],\cdots , \\
\lbrack k^{\prime }n/2+\frac{j-(n-3)}{2}+\frac{n-j}{n+1}+\frac{j-3}{2(n+1)}%
],[k^{\prime }n/2+\frac{j-(n-1)}{2}+\frac{n-j}{n+1}+\frac{j-1}{2(n+1)}]%
\end{array}%
\right\} .  \notag
\end{eqnarray}%
An even term forms a pair with an odd term when $\frac{j-(n-t)}{2(n+1)}=\frac{%
j-(r-1)}{2(n+1)},$ that is $r+t=n+1.$ We do the following approximation for
such a pair:
\begin{align}
\lbrack k^{\prime }n/2+\frac{j-(n-t)}{2}-\frac{j-(n-t)}{2(n+1)}]& [k^{\prime
}n/2+\frac{j-[(n-r)+1]}{2}+\frac{n-j}{n+1}+\frac{j-(r-1)}{2(n+1)}]\approx
\notag \\
& [k^{\prime }n/2+\frac{j-(n-t)}{2}][k^{\prime }n/2+\frac{j-[(n-r)+1]}{2}+%
\frac{n-j}{n+1}].
\end{align}%
After doing the approximation to all pairs, we have
\begin{equation}
\overline{a}_{kn,j} \approx \frac{(-1)^{k}}{2^{kn}(kn/2+\frac{j-1}{2}%
)!\Gamma (kn/2+\frac{j-1}{2}+\frac{n-j}{n+1}+1)}
=\frac{(-1)^{k}}{2^{kn}(kn/2+b_{j})!\Gamma (kn/2+b_{j}+\frac{n-j}{n+1}+1)}
\end{equation}%
with $a_{0,j}=\frac{1}{\Gamma (\frac{j-1}{2}+1)\Gamma (\frac{j-1}{2}+\frac{%
n-j}{n+1}+1)}. $

Now we have finished doing approximations to $\Psi _{j}(\phi _{0}).$ Before
the further transformation, we observe that
\begin{equation}
\sum\limits_{q=\pm 1,\text{ }\pm 3,\cdots ,\text{ }\pm (\frac{n}{2}-1)}\frac{%
1}{n/2}(e^{i\frac{\pi }{n}\ast q})^{m}=\left\{
\begin{array}{ll}
0, & \hbox{\text{ mod}$(m,n)\neq 0$;} \\
+1, & \hbox{\text{ mod}$(m/n,2)=0$;} \\
-1, & \hbox{\text{ mod}$(m/n,2)=1$,}%
\end{array}%
\right.  \label{piecewiseq}
\end{equation}%
(See S3 in Ref. \cite{SM}). Here $q$ has $\frac{n}{2}$ values. It is also
crucial to notice that $f\propto \kappa ^{\frac{1-n}{2}}\propto x^{\frac{1}{n%
}\frac{1-n}{2}}\propto \phi _{0}^{\frac{n}{n+1}\frac{1}{n}\frac{1-n}{2}%
}=\phi _{0}^{\frac{1-n}{2(n+1)}}=\phi _{0}^{\frac{2-(n+1)}{2(n+1)}},$ which
is a guide to approaching the asymptotic wave functions in Section \ref{asymptoticb}. Therefore, we transform $\Psi _{j}$ as follows:
\begin{eqnarray}
\Psi _{j} &=&\sum\limits_{k=0,1,2,\cdots }\frac{(-1)^{k}}{%
2^{kn}(kn/2+b_{j})!\Gamma (kn/2+b_{j}+v_{j}+1)}\phi _{0}^{kn+j\frac{n}{n+1}}
\notag \\
&=&\sum\limits_{k=0,1,2,\cdots }\frac{(-1)^{k}}{2^{kn}(kn/2+b_{j})!\Gamma
(kn/2+b_{j}+v_{j}+1)}\phi _{0}^{kn+2b_{j}+v_{j}+\frac{1}{n+1}}
\label{psijaa} \\
%&=&\phi _{0}^{\frac{1}{n+1}}\sum\limits_{k=0,1,2,\cdots }\frac{(-1)^{k}}{%
%2^{kn}(kn/2+b_{j})!\Gamma (kn/2+b_{j}+v_{j}+1)}\phi
%_{0}^{2(kn/2+b_{j})+v_{j}}  \notag \\
&=&\phi _{0}^{\frac{1}{n+1}}\sum\limits_{m=kn/2}\frac{(-1)^{k}}{%
2^{2m}(m+b_{j})!\Gamma (m+b_{j}+v_{j}+1)}\phi _{0}^{2(m+b_{j})+v_{j}}  \notag
\\
%&=&\phi _{0}^{\frac{1}{n+1}}\sum\limits_{m=kn/2}\frac{\sum\limits_{q=\pm 1,%
%\text{ }\pm 3,\cdots ,\text{ }\pm (\frac{n}{2}-1)}\frac{1}{n/2}(e^{i\frac{%
%\pi }{n}\ast q})^{2m}}{2^{2m}(m+b_{j})!\Gamma (m+b_{j}+v_{j}+1)}\phi
%_{0}^{2(m+b_{j})+v_{j}}  \notag \\
%&=&\frac{1}{n/2}\phi _{0}^{\frac{1}{n+1}}\sum\limits_{q=\pm 1,\text{ }\pm
%3,\cdots ,\text{ }\pm (\frac{n}{2}-1)}\sum\limits_{m=0}^{\infty }\frac{(e^{i%
%\frac{\pi }{n}\ast q})^{2m}}{2^{2m}(m+b_{j})!\Gamma (m+b_{j}+v_{j}+1)}\phi
%_{0}^{2(m+b_{j})+v_{j}}  \notag \\
&=&\frac{1}{n/2}2^{2b_{j}+v_{j}}\phi _{0}^{\frac{1}{n+1}}\sum\limits_{q=\pm
1,\text{ }\pm 3,\cdots ,\text{ }\pm (\frac{n}{2}-1)}(e^{i\frac{\pi }{n}\ast
q})^{-2b_{j}-v_{j}}\ast  \notag \\
&&\sum\limits_{m=0}^{\infty }\frac{(e^{i\frac{\pi }{n}\ast
q})^{2(m+b_{j})+v_{j}}}{2^{2(m+b_{j})+v_{j}}(m+b_{j})!\Gamma
(m+b_{j}+v_{j}+1)}\phi _{0}^{2(m+b_{j})+v_{j}}  \notag \\
&\overset{m^{\prime }=m+b_{j}}{=}&\frac{1}{n/2}2^{2b_{j}+v_{j}}\phi _{0}^{%
\frac{1}{n+1}}\sum\limits_{q=\pm 1,\text{ }\pm 3,\cdots ,\text{ }\pm (\frac{n%
}{2}-1)}(e^{i\frac{\pi }{n}\ast q})^{-2b_{j}-v_{j}}\ast  \notag \\
&&\sum\limits_{\text{ }m^{\prime }=0}^{\infty }\frac{(e^{i\frac{\pi }{n}\ast
q})^{2m^{\prime }+v_{j}}}{2^{2m^{\prime }+v_{j}}m^{\prime }!\Gamma
(m^{\prime }+v_{j}+1)}\phi _{0}^{2m^{\prime }+v_{j}}.  \notag
\end{eqnarray}

{{} In this part, we have expressed the $n$ independent wave functions by the integral $\phi
_{0}=\int_{0}^{x}\kappa _{0}dx^{\prime }=\frac{n}{n+1}[{} V
_{x=R_{E}}^{(1)}]^{1/n}x^{\frac{n+1}{n}}.$ Then, these wave functions can be represented by Bessel functions.}
\subsubsection{Bessel function representation}\label{bessel}
Without loss of generality, do the transformation $\Psi _{j}\rightarrow
\frac{n}{2}2^{-(2b_{j}+v_{j})}\Psi _{j}$ to remove a consant$.$ We know that
\begin{equation}
I_{\nu }(z)=\sum\limits_{m=0}^{\infty }\frac{1}{m!\Gamma (m+v+1)}(\frac{z}{2}%
)^{2m+\nu }.
\end{equation}
Now $\Psi _{j}$ can be represented by $n/2$ Bessel functions: ($m^{\prime
}\rightarrow m$ in Eq. (\ref{psijaa}) )
\begin{eqnarray}
\Psi _{j} &=&\phi _{0}^{\frac{1}{n+1}}\sum\limits_{q=\pm 1,\text{ }\pm
3,\cdots ,\text{ }\pm (\frac{n}{2}-1)}(e^{i\frac{\pi }{n}\ast
q})^{-2b_{j}-v_{j}}\sum\limits_{m=0}^{\infty }\frac{(e^{i\frac{\pi }{n}\ast
q})^{2m+v_{j}}}{2^{2m+v_{j}}m!\Gamma (m+v_{j}+1)}\phi _{0}^{2m+v_{j}}
\label{psiphi} \\
&=&\phi _{0}^{\frac{1}{n+1}}\sum\limits_{q=\pm 1,\text{ }\pm 3,\cdots ,\text{
}\pm (\frac{n}{2}-1)}(e^{i\frac{\pi }{n}\ast
q})^{-2b_{j}-v_{j}}\sum\limits_{m=0}^{\infty }\frac{1}{m!\Gamma (m+v_{j}+1)}(%
\frac{e^{i\frac{\pi }{n}\ast q}\phi _{0}}{2})^{2m+v_{j}}  \notag \\
&=&\phi _{0}^{\frac{1}{n+1}}\sum\limits_{q=\pm 1,\text{ }\pm 3,\cdots ,\text{
}\pm (\frac{n}{2}-1)}(e^{i\frac{\pi }{n}\ast q})^{-2b_{j}-v_{j}}I_{\nu
_{j}}(\phi _{q}),  \notag
\end{eqnarray}%
where $\phi _{q}=e^{i\frac{\pi }{n}\ast q}\phi _{0}.$

The asymptotic expansion of $I_{\nu }(z)$ in long distances is \cite%
{watson1995}
\begin{equation}
I_{\nu }(z)\rightarrow \frac{e^{z}}{(2\pi z)^{1/2}}[1+\ O(\frac{1}{z})]+%
\frac{e^{-z+(\nu +1/2)\pi i}}{(2\pi z)^{1/2}}[1+O(\frac{1}{z})].  \label{ivz}
\end{equation}%
As $-e^{i\frac{\pi }{n}\ast q}=e^{i\frac{\pi }{n}\ast (q\pm n)},$ the $%
e^{-z} $ modes correspond to $q=\pm (\frac{n}{2}+1),\pm (\frac{n}{2}%
+3),\cdots ,\pm (n-1),$ i.e. the $n/2$ dissipating wave components. Of
importance are the phases of the $e^{z}$ components, i.e. the exponentially
growing components. Substitute Eq. (\ref{ivz}) into Eq. (\ref{psiphi}), we
have
\begin{eqnarray}
\Psi _{j} &=&\phi _{0}^{\frac{2-(n+1)}{2(n+1)}}\sum\limits_{q=\pm 1,\text{ }%
\pm 3,\cdots ,\text{ }\pm (\frac{n}{2}-1)}(e^{i\frac{\pi }{n}\ast
q})^{-2b_{j}-v_{j}}e^{(i\frac{\pi }{n}\ast q)\frac{-1}{2}}e^{\phi
_{q}}+O(e^{-\phi _{q}})  \label{psikappa} \\
&=&\phi _{0}^{\frac{2-(n+1)}{2(n+1)}}\sum\limits_{q=\pm 1,\text{ }\pm
3,\cdots ,\text{ }\pm (\frac{n}{2}-1)}(e^{i\frac{\pi }{n}\ast
q})^{-2b_{j}-v_{j}-\frac{1}{2}}e^{\phi _{q}}+O(e^{-\phi _{q}}),  \notag
\end{eqnarray}%
where $(2\pi )^{-1/2}$ is ignored. We use $c_{j,q}=(e^{i\frac{\pi }{n}\ast
q})^{-2b_{j}-v_{j}-\frac{1}{2}}$ and $\Psi _{\kappa =e^{i\frac{\pi }{n}\ast
q}\kappa _{0}}=\phi _{0}^{\frac{2-(n+1)}{2(n+1)}}e^{\phi _{q}}$ in the
following. The asymptotic behavior of $\Psi _{j}$ evidently matches the
formula mentioned in Section \ref{asymptoticb}.

On the left side of $R_{E},$ we have
\begin{eqnarray}
{{}\frac{d^n\Psi(x)}{d x^n} } &=&[-{} V _{x=R_{E}}^{(1)}(x-R_{E})]\Psi ,  \notag
\\
\text{or }{{}\frac{d^n\Psi(x^{\prime })}{d x^{\prime n}} } &=&[{} V _{x^{\prime
}=0}^{(1)}x^{\prime }]\Psi   \label{sub2}
\end{eqnarray}%
with $x^{\prime }=R_{E}-x.$ For simplicity, $x^{\prime }\rightarrow x.$
Again, we apply the Frobenius method. Assume $\overline{\Psi }%
_{j}=\sum\limits_{m=0}^{\infty }a_{m}x^{m}.$ We get the recursive
relationship
\begin{equation}
\frac{a_{m+n+1}}{a_{m}}=\frac{{} V _{x=R_{E}}^{(1)}}{(m+n+1)(m+n+1-1)%
\cdots \lbrack m+n+1-(n-1)]}.
\end{equation}%
Since only a plus/minus symbol is different compared with Section \ref{ite1}, we
skip similar steps. Define $k_{0}=[E-{} V (x)]^{1/n}$ with arg$(k_{0})=0,$
{{} and the integral of $k_0$ from $R_E$ to a point on the left of $R_E$: $\overline{\phi }_{0}=\int_{0}^{x}k_{0}dx^{\prime\prime  }=\frac{n}{n+1}[{} V
_{x=R_{E}}^{(1)}]^{1/n}x^{\frac{n+1}{n}}.$} We have
\begin{eqnarray}  \label{phio}
\overline{\Psi }_{j} &=&\sum\limits_{k=0,1,2,\cdots }\frac{1}{%
2^{kn}(kn/2+b_{j})!\Gamma (kn/2+b_{j}+v_{j}+1)}\overline{\phi }_{0}^{kn+j%
\frac{n}{n+1}}  \label{k0} \\
&=&2^{2b_{j}+v_{j}}(+1)^{-2b_{j}/n}(+1)^{-v_{j}/n}\overline{\phi }_{0} ^{%
\frac{1}{n+1}}\sum\limits_{l^{\prime }=kn/2+b_{j}}\frac{(+1)^{(2l^{\prime
}+v_{j})/n}}{2^{2l^{\prime }+v_{j}}(l^{\prime }!)\Gamma (l^{\prime }+v_{j}+1)%
}\overline{\phi }_{0}^{2l^{\prime }+v_{j}}.  \notag
\end{eqnarray}

Now we need to use a substitution $\sum\limits_{\overline{q}=-\frac{n}{2}+2,%
\text{ }-\frac{n}{2}+4,\cdots ,\text{ }\frac{n}{2}}\frac{1}{n/2}(e^{i\frac{%
\pi }{n}\ast \overline{q}})^{m}=\left\{
\begin{array}{ll}
0, & \hbox{\text{ mod}$(m,n)\neq 0$;} \\
+1, & \hbox{\text{ mod}$(m,n)=0$.}%
\end{array}%
\right. $ Therefore,
\begin{eqnarray}
\overline{\Psi }_{j} &=&\overline{\phi }_{0}^{\frac{1}{n+1}}\sum\limits_{%
\overline{q}=-\frac{n}{2}+2,\text{ }-\frac{n}{2}+4,\cdots ,\text{ }\frac{n}{2%
}}(e^{i\frac{\pi }{n}\ast \overline{q}})^{-2b_{j}-v_{j}}\sum\limits_{m=0}^{%
\infty }\frac{(e^{i\frac{\pi }{n}\ast \overline{q}})^{2m+v_{j}}}{%
2^{2m+v_{j}}m!\Gamma (m+v_{j}+1)}\overline{\phi }_{0}^{2m+v_{j}}  \label{kp2}
\\
%&=&\overline{\phi }_{0}^{\frac{1}{n+1}}\sum\limits_{\overline{q}=-\frac{n}{2}%
%+2,\text{ }-\frac{n}{2}+4,\cdots ,\text{ }\frac{n}{2}}(e^{i\frac{\pi }{n}%
%\ast \overline{q}})^{-2b_{j}-v_{j}}\sum\limits_{m=0}^{\infty }\frac{1}{%
%m!\Gamma (m+v_{j}+1)}(\frac{e^{i\frac{\pi }{n}\ast \overline{q}}\overline{%
%\phi }_{0}}{2})^{2m+v_{j}}  \notag \\
&=&\overline{\phi }_{0}^{\frac{1}{n+1}}\sum\limits_{\overline{q}=-\frac{n}{2}%
+2,\text{ }-\frac{n}{2}+4,\cdots ,\text{ }\frac{n}{2}}(e^{i\frac{\pi }{n}%
\ast \overline{q}})^{-2b_{j}-v_{j}}I_{\nu _{j}}(\overline{\phi }_{\overline{q%
}}).  \notag
\end{eqnarray}

\begin{figure}[htb]
\centering
\includegraphics[width=9cm]{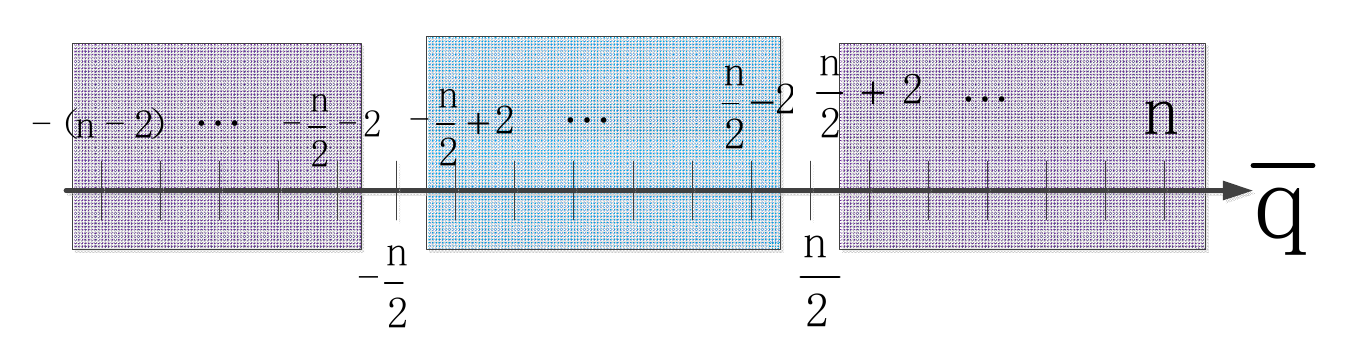}
\caption{The ranges of exponentially growing components (purple), free particle states, and dissipating components (indigo).}\label{plotq}
\end{figure}

For $\overline{\Psi }_{j},$ we have obtained the phase factor $\overline{c}%
_{j,\overline{q}}=(e^{i\frac{\pi }{n}\ast \overline{q}})^{-2b_{j}-v_{j}-%
\frac{1}{2}}$ and $\overline{\Psi }_{k=e^{i\frac{\pi }{n}\ast \overline{q}%
}k_{0}}=\phi _{0}^{\frac{2-(n+1)}{2(n+1)}}e^{\overline{\phi }_{\overline{q}%
}} $. Since the forms of $\overline{c}_{j,\overline{q}}$ and $c_{j,q}$ are
identical, we use $c_{j,\overline{q}}$ to replace $\overline{c}_{j,\overline{%
q}}.$ $\overline{q}$ is sometimes replaced by $q$ if clearness is not
damaged. There are $n/2-1$ exponentially growing components in $\overline{%
\Psi }_{k=e^{i\frac{\pi }{n}\ast \overline{q}}k_{0}}$: $\overline{q}=-\frac{n%
}{2}+2,$ $-\frac{n}{2}+4,\cdots ,$ $\frac{n}{2}-2;$ while $2$ components are the
``free particle states'': $\overline{q}=\pm
\frac{n}{2};$ and $n/2-1$ dissipating components: $\overline{q}=\pm (\frac{n}{%
2}+2),$ $\pm (\frac{n}{2}+4),\cdots ,$ $\pm (n-2),n$. Different ranges of $\overline{q}$ are shown in Fig. \ref{plotq}. The bound-state wave
functions on the right side of $R_{E}$ are the linear combinations of $\Psi
_{j}:$ $\Psi _{b}=\sum\limits_{j=0}^{n-1}\alpha _{j}\Psi _{j},$ and on the left side of $R_{E}:$ $\Psi _{b}=\sum\limits_{j=0}^{n-1}\overline{\alpha }%
_{j}\overline{\Psi }_{j}.$
\subsubsection{\protect\bigskip Continuous condition: $\overline{\protect%
\alpha }_{j}=(-1)^{j}\protect\alpha _{j}$}
Using the continuous condition near $R_{E},$ we have
\begin{equation}
\alpha _{j}\Psi _{j} \propto \alpha _{j}(x-R_{E})^{j}, \,  \label{sub3}
\overline{\alpha }_{j}\overline{\Psi }_{j} \propto \overline{\alpha }%
_{j}(R_{E}-x)^{j}, \,\,\text{and}  \,\,
\alpha _{j}(x-R_{E})^{j} =\overline{\alpha }_{j}(R_{E}-x)^{j},
\end{equation}%
where $x$ is the Cartesian coordinate without any substitutions. Therefore, $%
\overline{\alpha }_{j}=(-1)^{j}\alpha _{j}.$

On the right side of $R_{E},$ the relationship between $\Psi _{j}$ and $\Psi
_{\kappa }$ can be expressed as
\begin{equation}
\left(
\begin{array}{c}
\Psi _{0} \\
\vdots \\
\Psi _{j} \\
\vdots \\
\Psi _{n-1}%
\end{array}%
\right) =\left(
\begin{array}{ccc|c}
c_{j=0,q=-(\frac{n}{2}-1)} & \cdots & c_{j=0,q=+(\frac{n}{2}-1)} & \cdots \\
\vdots & \ddots & \vdots & \ddots \\
c_{j,q=-(\frac{n}{2}-1)} & \cdots & c_{j,q=+(\frac{n}{2}-1)} & \cdots \\
\vdots & \ddots & \vdots & \ddots \\
c_{j=n-1,q=-(\frac{n}{2}-1)} & \cdots & c_{j=n-1,q=+(\frac{n}{2}-1)} & \cdots%
\end{array}%
\right) \left(
\begin{array}{c}
\Psi _{\kappa _{1}} \\
\vdots \\
\Psi _{\kappa _{j-1}} \\
\vdots \\
\Psi _{\kappa _{n}}%
\end{array}%
\right) ,
\end{equation}%
where the last column denotes the elements of $n/2$ dissipating components.

On the left side of $R_{E},$ the relationship between $\overline{\Psi }_{j}$
and $\overline{\Psi }_{k}$ can be expressed as
\begin{equation}
\left(
\begin{array}{c}
\overline{\Psi }_{0} \\
\vdots \\
\overline{\Psi }_{j} \\
\vdots \\
\overline{\Psi }_{n-1}%
\end{array}%
\right) =\left(
\begin{array}{ccc|cc}
c_{j=0,q=-\frac{n}{2}+2} & \cdots & c_{j=0,q=\frac{n}{2}-2} & c_{j=0,q=+%
\frac{n}{2}} & \cdots \\
\vdots & \ddots & \vdots & \vdots & \ddots \\
c_{j,q=-\frac{n}{2}+2} & \cdots & c_{j,q=\frac{n}{2}-2} & c_{j,q=+\frac{n}{2}%
} & \cdots \\
\vdots & \ddots & \vdots & \vdots & \ddots \\
c_{j=n-1,q=-\frac{n}{2}+2} & \cdots & c_{j=n-1,q=\frac{n}{2}-2} &
c_{j=n-1,q=+\frac{n}{2}} & \cdots%
\end{array}%
\right) \left(
\begin{array}{c}
\overline{\Psi }_{k_{1}} \\
\vdots \\
\overline{\Psi }_{k_{j-1}} \\
\vdots \\
\overline{\Psi }_{k_{n}}%
\end{array}%
\right) ,
\end{equation}%
where the last two columns on the right side of the vertical line represent the
elements of $n/2-1$ dissipating components and $2$ free particle components.

\subsubsection{$\protect\alpha _{j}=\protect\alpha _{n-1-j}$}

For the bound states, we use the following hypothesis.

\begin{hyp}
All exponentially growing components are negligible.
\end{hyp}

This hypothesis includes the exponential growing components from $R_{E}$ to
the right infinity, from $L_{E}$ to the left infinity, from $L_{E}$ to $%
R_{E} $, and from $R_{E}$ to $L_{E}. $

Let us see the case near $R_{E}$ first. To make these exponential growing
components vanish, the right side of $R_{E}$ gives $n/2$ equations, while the
left side of $R_{E}$ gives $n/2-1$ equations. Implementing the continuous
condition at $R_{E}$\ point $\overline{\alpha }_{j}=(-1)^{j}\alpha _{j},$ we
have
\begin{equation}
\left(
\begin{array}{c}
+\alpha _{0} \\
\vdots \\
\left( -1\right) ^{j(q+1)}\alpha _{j} \\
\vdots \\
\left( -1\right) ^{q+1}\alpha _{n-1}%
\end{array}%
\right) ^{T}\left(
\begin{array}{ccc|ccc}
c_{j=0,q=-(\frac{n}{2}-1)} & \cdots & c_{j=0,q=+(\frac{n}{2}-1)} & c_{j=0,q=-%
\frac{n}{2}+2} & \cdots & c_{j=0,q=\frac{n}{2}-2} \\
\vdots & \ddots & \vdots & \vdots & \ddots & \vdots \\
c_{j,q=-(\frac{n}{2}-1)} & \cdots & c_{j,q=+(\frac{n}{2}-1)} & c_{j,q=-\frac{%
n}{2}+2} & \cdots & c_{j,q=\frac{n}{2}-2} \\
\vdots & \ddots & \vdots & \vdots & \ddots & \vdots \\
c_{j=n-1,q=-(\frac{n}{2}-1)} & \cdots & c_{j=n-1,q=+(\frac{n}{2}-1)} &
c_{j=n-1,q=-\frac{n}{2}+2} & \cdots & c_{j=n-1,q=\frac{n}{2}-2}%
\end{array}%
\right) \left(
\begin{array}{c}
\Psi _{\kappa _{1}} \\
\vdots \\
\Psi _{\kappa _{n/2}} \\
\overline{\Psi }_{k_{1}} \\
\vdots \\
\overline{\Psi }_{k_{n/2-1}}%
\end{array}%
\right) =0,
\end{equation}%
which is equivalent to:
\begin{equation}
\left(
\begin{array}{ccccc}
c_{q=-(\frac{n}{2}-1),j=0} & \cdots & c_{q=-(\frac{n}{2}-1),j} & \cdots &
c_{q=-(\frac{n}{2}-1),j=n-1} \\
\vdots & \ddots & \vdots & \ddots & \vdots \\
c_{q=+(\frac{n}{2}-1),j=0} & \cdots & c_{q=+(\frac{n}{2}-1),j} & \cdots &
c_{q=+(\frac{n}{2}-1),j=n-1} \\
c_{q=-\frac{n}{2}+2,j=0} & \ddots & c_{q=-\frac{n}{2}+2,j} & \cdots & c_{q=-%
\frac{n}{2}+2,j=n-1} \\
\vdots & \ddots & \vdots & \ddots & \vdots \\
c_{q=\frac{n}{2}-2,j=0} & \cdots & c_{q=\frac{n}{2}-2,j} & \cdots & c_{q=%
\frac{n}{2}-2,j=n-1}%
\end{array}%
\right) \left(
\begin{array}{c}
+\alpha _{0} \\
\vdots \\
\vdots \\
\left( -1\right) ^{j(q+1)}\alpha _{j} \\
\vdots \\
\vdots \\
\left( -1\right) ^{q+1}\alpha _{n-1}%
\end{array}%
\right) =0,
\end{equation}%
i.e.,
\begin{equation}
\left(
\begin{array}{ccccc}
+c_{q=-(\frac{n}{2}-1),j=0} & \cdots & \left( -1\right) ^{j(q+1)}c_{q=-(%
\frac{n}{2}-1),j} & \cdots & \left( -1\right) ^{q+1}c_{q=-(\frac{n}{2}%
-1),j=n-1} \\
\vdots & \ddots & \vdots & \ddots & \vdots \\
+c_{q=+(\frac{n}{2}-1),j=0} & \cdots & \left( -1\right) ^{j(q+1)}c_{q=+(%
\frac{n}{2}-1),j} & \cdots & \left( -1\right) ^{q+1}c_{q=+(\frac{n}{2}%
-1),j=n-1} \\
+c_{q=-\frac{n}{2}+2,j=0} & \ddots & \left( -1\right) ^{j(q+1)}c_{q=-\frac{n%
}{2}+2,j} & \cdots & \left( -1\right) ^{q+1}c_{q=-\frac{n}{2}+2,j=n-1} \\
\vdots & \ddots & \vdots & \ddots & \vdots \\
+c_{q=\frac{n}{2}-2,j=0} & \cdots & \left( -1\right) ^{j(q+1)}c_{q=\frac{n}{2%
}-2,j} & \cdots & \left( -1\right) ^{q+1}c_{q=\frac{n}{2}-2,j=n-1}%
\end{array}%
\right) \left(
\begin{array}{c}
\alpha _{0} \\
\vdots \\
\alpha _{j} \\
\vdots \\
\alpha _{n-1}%
\end{array}%
\right) =0.
\end{equation}

We separate $\alpha _{0}$ from the left side, and get
\begin{equation}
\left(
\begin{array}{ccccc}
\left( -1\right) ^{(q+1)}c_{q=-\frac{n}{2}+1,j=1} & \cdots & \left(
-1\right) ^{j(q+1)}c_{q=-\frac{n}{2}+1,j} & \cdots & \left( -1\right)
^{q+1}c_{q=-\frac{n}{2}+1,j=n-1} \\
\left( -1\right) ^{(q+1)}c_{q=-\frac{n}{2}+2,j=1} & \ddots & \left(
-1\right) ^{j(q+1)}c_{q=-\frac{n}{2}+2,j} & \cdots & \left( -1\right)
^{q+1}c_{q=-\frac{n}{2}+2,j=n-1} \\
\vdots & \ddots & \vdots & \ddots & \vdots \\
\left( -1\right) ^{(q+1)}c_{q=\frac{n}{2}-2,j=1} & \cdots & \left( -1\right)
^{j(q+1)}c_{q=\frac{n}{2}-2,j} & \cdots & \left( -1\right) ^{q+1}c_{q=\frac{n%
}{2}-2,j=n-1} \\
\left( -1\right) ^{(q+1)}c_{q=\frac{n}{2}-1,j=1} & \cdots & \left( -1\right)
^{j(q+1)}c_{q=\frac{n}{2}-1,j} & \cdots & \left( -1\right) ^{q+1}c_{q=\frac{n%
}{2}-1,j=n-1}%
\end{array}%
\right) \left(
\begin{array}{c}
\alpha _{1} \\
\vdots \\
\alpha _{j} \\
\vdots \\
\alpha _{n-1}%
\end{array}%
\right) =-\left(
\begin{array}{c}
c_{q=-\frac{n}{2}+1,j=0} \\
c_{q=-\frac{n}{2}+2,j=0} \\
\vdots \\
c_{q=\frac{n}{2}-2,j=0} \\
c_{q=\frac{n}{2}-1,j=0}%
\end{array}%
\right) \alpha _{0}.
\end{equation}
Let $A=\left(
\begin{array}{ccccc}
\left( -1\right) ^{(q+1)}c_{q=-\frac{n}{2}+1,j=1} & \cdots & \left(
-1\right) ^{j(q+1)}c_{q=-\frac{n}{2}+1,j} & \cdots & \left( -1\right)
^{q+1}c_{q=-\frac{n}{2}+1,j=n-1} \\
\left( -1\right) ^{(q+1)}c_{q=-\frac{n}{2}+2,j=1} & \cdots & \left(
-1\right) ^{j(q+1)}c_{q=-\frac{n}{2}+2,j} & \cdots & \left( -1\right)
^{q+1}c_{q=-\frac{n}{2}+2,j=n-1} \\
\vdots & \ddots & \vdots & \ddots & \vdots \\
-c_{q=0,j=1} & \cdots & \left( -1\right) ^{j}c_{q=0,j} & \cdots &
-c_{q=0,j=n-1} \\
\vdots & \ddots & \vdots & \ddots & \vdots \\
\left( -1\right) ^{(q+1)}c_{q=\frac{n}{2}-2,j=1} & \cdots & \left( -1\right)
^{j(q+1)}c_{q=\frac{n}{2}-2,j} & \cdots & \left( -1\right) ^{q+1}c_{q=\frac{n%
}{2}-2,j=n-1} \\
\left( -1\right) ^{(q+1)}c_{q=\frac{n}{2}-1,j=1} & \cdots & \left( -1\right)
^{j(q+1)}c_{q=\frac{n}{2}-1,j} & \cdots & \left( -1\right) ^{q+1}c_{q=\frac{n%
}{2}-1,j=n-1}%
\end{array}%
\right) ,$ and $b=-\left(
\begin{array}{c}
c_{q=-\frac{n}{2}+1,j=0} ,
c_{q=-\frac{n}{2}+2,j=0},
\cdots,
c_{q=\frac{n}{2}-2,j=0} ,
c_{q=\frac{n}{2}-1,j=0}%
\end{array}%
\right)^T .$ Using Cramer's rule, $\alpha _{\underline{j}}$ is given by
\begin{equation}
\alpha _{\underline{j}}=\frac{\det (A_{\underline{j}})}{\det (A)}\alpha _{0},
\label{alphaj}
\end{equation}%
where $A_{\underline{j}}$ is obtained by replacing $\underline{j}$th column
of $A$ with $b.$ Now we prove $\alpha _{\underline{j}}=\alpha _{n-\underline{%
j}-1}.$ In general, $\det (A)\neq 0.$ Remark \ref{remarkdet} will
further discuss this point. It suffices to prove $\det (A_{\underline{j}%
})=\det (A_{n-\underline{j}-1}).$ Use $A_{\underline{j}}c_{_{j^{\prime }}}$
to denote the column of $A_{\underline{j}}$ with $j=j^{\prime },$ and use $%
A_{\underline{j}}r_{_{q^{\prime }}}$ to denote the row of $A_{\underline{j}}$
with $q=q^{\prime }.$ Accordingly, $A_{\underline{j}}=(A_{\underline{j}%
}c_{_{j^{\prime }=1}},\cdots ,A_{\underline{j}}c_{_{j^{\prime }=\underline{j}%
-1}},A_{\underline{j}}c_{_{j^{\prime }=0}},A_{\underline{j}}c_{_{j^{\prime }=%
\underline{j}+1}},\cdots ,A_{\underline{j}}c_{_{j^{\prime }=n-1}}),$ and $A_{%
\underline{j}}=(A_{\underline{j}}r_{_{q^{\prime }=-\frac{n}{2}+1}},\cdots
,A_{\underline{j}}r_{_{q^{\prime }=0}},\cdots ,A_{\underline{j}%
}r_{_{q^{\prime }=\frac{n}{2}-1}})^{T}.$ For $A_{\underline{j}},$ we move
the $c_{_{j^{\prime }=0}}$ column to the first column; for $A_{n-\underline{j%
}-1},$ we reverse the rows first, then reverse the columns, and move the $%
c_{_{j^{\prime }=0}}$ column to the last column. See below,
\begin{eqnarray}
\text{column operation}&\text{:}&\det (A_{\underline{j}})=\det (A_{%
\underline{j}}c_{_{j^{\prime }=1}},\cdots ,A_{\underline{j}}c_{_{j^{\prime
}=0}},\cdots ,A_{\underline{j}}c_{_{j^{\prime }=n-1}})=  \notag \\
&&\left( -1\right) ^{\underline{j}-1}\det (A_{\underline{j}}c_{_{j^{\prime
}=0}},A_{\underline{j}}c_{_{j^{\prime }=1}},\cdots ,A_{\underline{j}%
}c_{_{j^{\prime }=\underline{j}-1}},A_{\underline{j}}c_{_{j^{\prime }=%
\underline{j}+1}},\cdots ,A_{\underline{j}}c_{_{j^{\prime }=n-1}});
\end{eqnarray}%
\begin{eqnarray}
\text{row operation}\text{: } &&\det (A_{n-\underline{j}-1})=\det (A_{n-%
\underline{j}-1}r_{_{q^{\prime }=-\frac{n}{2}+1}},\cdots ,A_{n-\underline{j}%
-1}r_{_{q^{\prime }=0}},\cdots ,A_{n-\underline{j}-1}r_{_{q^{\prime }=\frac{n%
}{2}-1}})^{T}  \notag \\
&=&\left( -1\right) ^{\frac{n-2}{2}}\det (A_{n-\underline{j}%
-1}r_{_{_{q^{\prime }=\frac{n}{2}-1}}},\cdots ,A_{n-\underline{j}%
-1}r_{_{q^{\prime }=0}},\cdots ,A_{n-\underline{j}-1}r_{_{q^{\prime }=-\frac{%
n}{2}+1}})^{T} \\
&=&\left( -1\right) \det (A_{n-\underline{j}-1}r_{_{_{q^{\prime }=\frac{n}{2}%
-1}}},\cdots ,A_{n-\underline{j}-1}r_{_{q^{\prime }=0}},\cdots ,A_{n-%
\underline{j}-1}r_{_{q^{\prime }=-\frac{n}{2}+1}})^{T};  \notag
\end{eqnarray}%
and
\begin{align}\notag
&\text{column operation}\,\text{(after the row operation)}\text{:}\\
&\det (A_{n-\underline{j}-1}) \notag\\
&= -1\det (A_{n-\underline{j}-1}c_{_{j^{\prime
}=1}},\cdots ,A_{n-\underline{j}-1}c_{_{j^{\prime }=n-\underline{j}-2}},A_{n-%
\underline{j}-1}c_{_{j^{\prime }=0}},A_{n-\underline{j}-1}c_{_{j^{\prime }=n-%
\underline{j}}},   \cdots ,A_{n-\underline{j}-1}c_{_{j^{\prime }=n-1}}) \\
&= \left( -1\right) ^{2}\det (A_{n-\underline{j}-1}c_{_{j^{\prime
}=n-1}},\cdots ,A_{n-\underline{j}-1}c_{_{j^{\prime }=n-\underline{j}}},A_{n-%
\underline{j}-1}c_{_{j^{\prime }=0}},A_{n-\underline{j}-1}c_{_{j^{\prime }=n-%
\underline{j}-2}},  \cdots ,A_{n-\underline{j}-1}c_{_{j^{\prime }=1}})  \notag \\
&= \left( -1\right) ^{\underline{j}+2}\det (A_{n-\underline{j}%
-1}c_{_{j^{\prime }=n-1}},\cdots ,A_{n-\underline{j}-1}c_{_{j^{\prime }=n-%
\underline{j}}},A_{n-\underline{j}-1}c_{_{j^{\prime }=n-\underline{j}%
-2}},\cdots ,  A_{n-\underline{j}-1}c_{_{j^{\prime }=1}},A_{n-\underline{j}%
-1}c_{_{j^{\prime }=0}}).  \notag
\end{align}
Now, we only need to prove
\begin{multline}
\det (A_{\underline{j}}c_{_{j^{\prime }=0}},A_{\underline{j}}c_{_{j^{\prime
}=1}},\cdots ,A_{\underline{j}}c_{_{j^{\prime }=\underline{j}-1}},A_{%
\underline{j}}c_{_{j^{\prime }=\underline{j}+1}},\cdots ,A_{\underline{j}%
}c_{_{j^{\prime }=n-1}})= \\
-1\det (A_{n-\underline{j}-1}c_{_{j^{\prime }=n-1}},\cdots ,A_{n-\underline{j%
}-1}c_{_{j^{\prime }=n-\underline{j}}},A_{n-\underline{j}-1}c_{_{j^{\prime
}=n-\underline{j}-2}},\cdots ,A_{n-\underline{j}-1}c_{_{j^{\prime }=1}},A_{n-%
\underline{j}-1}c_{_{j^{\prime }=0}}).
\end{multline}

Remove a minus symbol in the column $b$ in both $A_{\underline{j}}$ and $%
A_{n-\underline{j}-1}.$ Without loss of rigorosity, we always use the same
notations $\det (A_{\underline{j}})$ and $\det (A_{n-\underline{j}-1})$ by
removing a common factor in the following. Currently, after the row and
column operations, we have the following form %$\rlap{Q}---$%
\begin{equation}
A_{\underline{j}}\rightarrow \left(
\begin{array}{cccccccc}
q\ \ \backslash \ \ j & 0 & 1 & \cdots & {\underline{j}}\llap{$\diagdown$} &
\cdots & n-2 & n-1 \\
-\frac{n}{2}+1 &  &  &  & \diagdown &  &  &  \\
\vdots &  &  &  & \diagdown &  &  &  \\
$\rlap{}--${0}\llap{--} & $\rlap{}--$ $\rlap{}--$ & $\rlap{}--$ $\rlap{}--$
& $\rlap{}--$ $\rlap{}--$ & \diagdown & $\rlap{}--$$\rlap{}--$ & $\rlap{}--$$%
\rlap{}--$ & $\rlap{}--$$\rlap{}--$ \\
\vdots &  &  &  & \diagdown &  &  &  \\
\frac{n}{2}-1 &  &  &  & \diagdown &  &  &
\end{array}%
\right) ,
\end{equation}%
and
\begin{equation}
A_{n-\underline{j}-1}\rightarrow \left(
\begin{array}{cccccccc}
q\ \ \backslash \ \ j & n-1 & n-2 & \cdots & {n-1\llap{$\diagdown$}-%
\underline{j}} & \cdots & 1 & 0 \\
\frac{n}{2}-1 &  &  &  & \diagdown &  &  &  \\
\vdots &  &  &  & \diagdown &  &  &  \\
$\rlap{}--${0}\llap{--} & $\rlap{}--$$\rlap{}--$ & $\rlap{}--$$\rlap{}--$ & $%
\rlap{}--$$\rlap{}--$ & \diagdown & $\rlap{}--$$\rlap{}--$ & $\rlap{}--$$%
\rlap{}--$ & $\rlap{}--$$\rlap{}--$ \\
\vdots &  &  &  & \diagdown &  &  &  \\
-\frac{n}{2}+1 &  &  &  & \diagdown &  &  &
\end{array}%
\right),
\end{equation}
{{} where the notations ``$ {\underline{j}}\llap{$\diagdown$} $'' and `` $ {n-1\llap{$\diagdown$}-%
\underline{j}}$ '' mean that  the column $A_{\underline{j}}c_{_{j^{\prime }=\underline{j}}}$ and $A_{n-\underline{j}-1}c_{_{j^{\prime
}=n-\underline{j}-1}}$do not exist respectively, and the notation ``$\rlap{}-{0}\llap{--}\rlap{}-$''  means that we use the $r_{_{q=0}}$ row to do the Laplace expansion.}

We notice that $q$'s are symmetric about $0$. To calculate the determinant,
we use the $r_{_{q=0}}$ row to do the Laplace expansion. Later we will see
the benefits.
\begin{eqnarray}
\det (A_{\underline{j}}) &=&\sum\limits_{j=0,1,\cdots ,n-1,\neq _{\underline{%
j}}}(-1)^{\overline{q}+\overline{j}}\left( -1\right) ^{j}c_{q=0,j}M_{%
\overline{q}\overline{j}}, \\
\det (A_{n-\underline{j}-1}) &=&-1\sum\limits_{j=n-1,n-2,\cdots ,0,\neq
_{n-1-\underline{j}}}(-1)^{\overline{q}+\overline{j}}\left( -1\right)
^{j}c_{q=0,j}M_{\overline{q}\overline{j}},  \notag
\end{eqnarray}%
where $\overline{q}$th and $\overline{j}$th denote the row and column
numbers, respectively. $\left( -1\right) ^{j}c_{q=0,j}$ is the entry in the $%
(\overline{q},\overline{j})$ position, and $(-1)^{\overline{q}+\overline{j}%
}M_{\overline{q}\overline{j}}$ the cofactor. We now prove that for a
specific $(\overline{q},\overline{j}),$
\begin{equation}
(-1)^{\overline{q}+\overline{j}}\left( -1\right) ^{j_{1}}c_{q=0,j_{1}}M_{1%
\overline{q}\overline{j}}=-1(-1)^{\overline{q}+\overline{j}}\left( -1\right)
^{j_{2}}c_{q=0,j_{2}}M_{2\overline{q}\overline{j}},
\end{equation}%
where $j_{1}$ and $j_{2}$ are $j$'s in $A_{\underline{j}}$ and $A_{n-%
\underline{j}-1},$ respectively. Obviously, $j_{1}+j_{2}=n-1.$

In the present form of $A_{\underline{j}}$ and $A_{n-\underline{j}-1},$ we
notice that two symmetries exist: $q$'s are symmetric about $0$, and $%
j_{1}+j_{2}=n-1$ at $\overline{j}$th column. Besides, $c_{q=0,j}=(e^{i\frac{%
\pi }{n}\ast 0})^{-2b_{j}-v_{j}-\frac{1}{2}}=1.$ Now we need to prove
\begin{equation}
M_{1\overline{q}\overline{j}}=M_{2\overline{q}\overline{j}}.  \label{cofa2}
\end{equation}

Remember that $M_{1\overline{q}\overline{j}}$ and $M_{2\overline{q}\overline{%
j}}$ are the determinants of the matrices formed by deleting $\overline{q}$%
th row and $\overline{j}$th column in $A_{\underline{j}}$ and $A_{n-%
\underline{j}-1},$ respectively. Use the notations $M_{1\overline{q}%
\overline{j}}=\det (A_{\underline{j}}^{\overline{q}\overline{j}})\ $and $M_{2%
\overline{q}\overline{j}}=\det (A_{n-\underline{j}-1}^{\overline{q}\overline{%
j}}).$ The entries are $\left( -1\right) ^{j(q+1)}c_{q,j}.$ Since $\nu _{j}=j%
\frac{n}{n+1}-2b_{j}-\frac{1}{n+1},$ we have $\left( -1\right)
^{j(q+1)}c_{q,j}=\left( -1\right) ^{j(q+1)}(e^{i\frac{\pi }{n}\ast
q})^{-2b_{j}-v_{j}-\frac{1}{2}}=\left( -1\right) ^{j(q+1)}(e^{i\frac{\pi }{n}%
\ast q})^{-j\frac{n}{n+1}+\frac{1}{n+1}-\frac{1}{2}}.$ According to the
Leibniz formula,
\begin{eqnarray}
\det (A_{\underline{j}}^{\overline{q}\overline{j}}) &=&\sum\limits_{\sigma
_{1}\in S_{n-2}}[\text{sgn}(\sigma _{1})\prod\limits_{\overline{q}_{1}=1}^{%
\overline{q}_{1}=n-2}\left( -1\right) ^{j_{1\sigma _{\overline{q}_{1}}}(q_{1%
\overline{q}_{1}}+1)}c_{\overline{q}_{1},\sigma _{\overline{q}_{1}}}^{\prime
}],\text{ and} \\
\det (A_{n-\underline{j}-1}^{\overline{q}\overline{j}})
&=&\sum\limits_{\sigma _{2}\in S_{n-2}}[\text{sgn}(\sigma _{2})\prod\limits_{%
\overline{q}_{2}=1}^{\overline{q}_{2}=n-2}\left( -1\right) ^{j_{2\sigma _{%
\overline{q}_{2}}}(q_{2\overline{q}_{2}}+1)}c_{\overline{q}_{2},\sigma _{%
\overline{q}_{2}}}^{\prime }].  \notag
\end{eqnarray}%
$j_{1\sigma _{\overline{q}_{1}}}(j_{2\sigma _{\overline{q}_{2}}})$ means the
value $j$ at the $\sigma _{\overline{q}_{1}}(\sigma _{\overline{q}_{2}})$th
column in $A_{\underline{j}}^{\overline{q}\overline{j}}(A_{n-\underline{j}%
-1}^{\overline{q}\overline{j}})$, $q_{1\overline{q}_{1}}(q_{2\overline{q}%
_{2}})$ is the value $q$ at the $\overline{q}_{1}(\overline{q}_{2})$th row,
and $c_{\overline{q},\sigma _{\overline{q}}}^{\prime }=c_{q_{\overline{q}%
},j_{\sigma _{_{\overline{q}}}}}$. In the following, $q\neq 0,$ and for
simplicity, define $\sigma _{q_{\overline{q}}}=\sigma _{\overline{q}}.$
Consider each $\sigma _{1}=\sigma _{2}=\sigma $ permutation term:
\begin{eqnarray}
\prod\limits_{\overline{q}_{1}=1}^{\overline{q}_{1}=n-2}\left( -1\right)
^{j_{1\sigma _{\overline{q}_{1}}}(q_{1\overline{q}_{1}}+1)}c_{\overline{q}%
_{1},\sigma _{\overline{q}_{1}}}^{\prime } &=&\prod\limits_{q_{1}=-\frac{n}{2%
}+1}^{q_{1}=\frac{n}{2}-1}\left( -1\right) ^{j_{1\sigma
_{q_{1}}}(q_{1}+1)}c_{q_{1},j_{1\sigma _{q_{1}}}}^{{}} \\
&=&\prod\limits_{q_{1}=-\frac{n}{2}+1}^{q_{1}=\frac{n}{2}-1}\left( -1\right)
^{j_{1\sigma _{q_{1}}}(q_{1}+1)}\prod\limits_{q_{1}=-\frac{n}{2}+1}^{q_{1}=%
\frac{n}{2}-1}c_{q_{1},j_{1\sigma _{q_{1}}}}^{{}},  \,\,\text{and}\notag \\
\prod\limits_{\overline{q}_{2}=1}^{\overline{q}_{2}=n-2}\left( -1\right)
^{j_{2\sigma _{\overline{q}_{2}}}(q_{2\overline{q}_{2}}+1)}c_{\overline{q}%
_{2},\sigma _{\overline{q}_{2}}}^{\prime } &=&\prod\limits_{q_{2}=\frac{n}{2}%
-1}^{q_{2}=-\frac{n}{2}+1}\left( -1\right) ^{j_{2\sigma
_{q_{2}}}(q_{2}+1)}c_{q_{2},j_{2}\sigma _{q_{2}}}^{{}}  \notag \\
&=&\prod\limits_{q_{2}=\frac{n}{2}-1}^{q_{2}=-\frac{n}{2}+1}\left( -1\right)
^{j_{2\sigma _{q_{2}}}(q_{2}+1)}\prod\limits_{q_{2}=\frac{n}{2}-1}^{q_{2}=-%
\frac{n}{2}+1}c_{q_{2},j_{2\sigma _{q_{2}}}}^{{}}  \notag \\
&=&\prod\limits_{q_{2}=\frac{n}{2}-1}^{q_{2}=-\frac{n}{2}+1}\left( -1\right)
^{(n-1-j_{1\sigma _{q_{2}}})(q_{2}+1)}\prod\limits_{q_{2}=\frac{n}{2}%
-1}^{q_{2}=-\frac{n}{2}+1}c_{q_{2},j_{2\sigma _{q_{2}}}}^{{}}.  \notag
\end{eqnarray}

Handle two $\prod $ terms separately,
\begin{eqnarray}
&&\prod\limits_{q_{2}=\frac{n}{2}-1}^{q_{2}=-\frac{n}{2}+1}\left( -1\right)
^{(n-1-j_{1\sigma _{q_{2}}})(q_{2}+1)}\\
&&=\prod\limits_{q_{2}=\frac{n}{2}%
-1}^{q_{2}=-\frac{n}{2}+1}\left( -1\right) ^{(n-1)(q_{2}+1)-j_{1\sigma
_{q_{2}}}(q_{2}+1)}
=\prod\limits_{q_{2}=\frac{n}{2}-1}^{q_{2}=-\frac{n}{2}+1}\left( -1\right)
^{(n-1)(q_{2}+1)-j_{1\sigma _{q_{2}}}(q_{2}+1)}  \notag \\
&&=\left( -1\right) ^{\sum\limits_{q_{2}=\frac{n}{2}-1}^{q_{2}=-\frac{n}{2}%
+1}(n-1)(q_{2}+1)}\prod\limits_{q_{2}=\frac{n}{2}-1}^{q_{2}=-\frac{n}{2}%
+1}\left( -1\right) ^{-j_{1\sigma _{q_{2}}}(q_{2}+1)}  \notag
=\prod\limits_{q_{2}=\frac{n}{2}-1}^{q_{2}=-\frac{n}{2}+1}\left( -1\right)
^{-j_{1\sigma _{q_{2}}}(q_{2}+1)}  \notag \\
&&=\prod\limits_{q_{2}=\frac{n}{2}-1}^{q_{2}=-\frac{n}{2}+1}\left( -1\right)
^{j_{1\sigma _{q_{2}}}(-q_{2}-1)}  \notag
=\prod\limits_{q_{1}=-\frac{n}{2}+1}^{q_{1}=\frac{n}{2}-1}\left( -1\right)
^{j_{1\sigma _{q_{1}}}(q_{1}-1)}  \notag
=\prod\limits_{q_{1}=-\frac{n}{2}+1}^{q_{1}=\frac{n}{2}-1}\left( -1\right)
^{j_{1\sigma _{q_{1}}}(q_{1}+1)}.  \notag
\end{eqnarray}%
Note that $\sigma _{q_{2=q}}=\sigma _{q_{1=-q}}.$ Thus, $\prod%
\limits_{q_{1}=-\frac{n}{2}+1}^{q_{1}=\frac{n}{2}-1}\left( -1\right)
^{j_{1\sigma _{q_{1}}}(q_{1}+1)}=\prod\limits_{q_{2}=\frac{n}{2}-1}^{q_{2}=-%
\frac{n}{2}+1}\left( -1\right) ^{(n-1-j_{1\sigma _{q_{2}}})(q_{2}+1)}.$ For
another $\prod $,
\begin{eqnarray}
\prod\limits_{q_{2}=\frac{n}{2}-1}^{q_{2}=-\frac{n}{2}+1}c_{q_{2},j_{2\sigma
_{q_{2}}}}^{{}} &=&\prod\limits_{q_{2}=\frac{n}{2}-1}^{q_{2}=-\frac{n}{2}%
+1}(e^{i\frac{\pi }{n}\ast q_{2}})^{-j_{2\sigma _{q_{2}}}\frac{n}{n+1}+\frac{%
1}{n+1}-\frac{1}{2}} \\
&=&\prod\limits_{q_{2}=\frac{n}{2}-1}^{q_{2}=-\frac{n}{2}+1}(e^{i\frac{\pi }{%
n}\ast q_{2}})^{\frac{1}{n+1}-\frac{1}{2}}\prod\limits_{q_{2}=\frac{n}{2}%
-1}^{q_{2}=-\frac{n}{2}+1}e^{-i\frac{\pi }{n}\frac{n}{n+1}q_{2}j_{2\sigma
_{q_{2}}}},  \notag
\end{eqnarray}
\begin{equation}
\prod\limits_{q_{2}=\frac{n}{2}-1}^{q_{2}=-\frac{n}{2}+1}(e^{i\frac{\pi }{n}%
\ast q_{2}})^{\frac{1}{n+1}-\frac{1}{2}}=\prod\limits_{q_{1}=-\frac{n}{2}%
+1}^{q_{1}=\frac{n}{2}-1}(e^{i\frac{\pi }{n}\ast q_{1}})^{\frac{1}{n+1}-%
\frac{1}{2}},
\end{equation}%
and
\begin{eqnarray}
&&\prod\limits_{q_{2}=\frac{n}{2}-1}^{q_{2}=-\frac{n}{2}+1}e^{-i\frac{\pi }{n}%
\frac{n}{n+1}q_{2}j_{2\sigma _{q_{2}}}}\\
&&=e^{-i\frac{\pi }{n}\frac{n}{n+1}%
\sum\limits_{q_{2}=\frac{n}{2}-1}^{q_{2}=-\frac{n}{2}+1}q_{2}j_{2\sigma
_{q_{2}}}} =e^{-i\frac{\pi }{n}\frac{n}{n+1}\sum\limits_{q_{2}=\frac{n}{2}%
-1}^{q_{2}=1}q_{2}(j_{2\sigma _{q_{2}}}-j_{2\sigma _{-q_{2}}})}  \notag \\
&&=e^{-i\frac{\pi }{n}\frac{n}{n+1}\sum\limits_{q_{2}=\frac{n}{2}%
-1}^{q_{2}=1}q_{2}[(n-1-j_{1\sigma _{q_{2}}})-(n-1-j_{1\sigma -_{q_{2}}})]}
\notag
=e^{-i\frac{\pi }{n}\frac{n}{n+1}\sum\limits_{q_{1}=-\frac{n}{2}%
+1}^{q_{1}=-1}-q_{1}[(n-1-j_{1\sigma _{q_{1}}})-(n-1-j_{1\sigma -_{q_{1}}})]}
\notag \\
&&=e^{-i\frac{\pi }{n}\frac{n}{n+1}\sum\limits_{q_{1}=-\frac{n}{2}%
+1}^{q_{1}=-1}q_{1}(j_{1\sigma _{q_{1}}}-j_{1\sigma -_{q_{1}}})}  \notag
=e^{-i\frac{\pi }{n}\frac{n}{n+1}\sum\limits_{q_{1}=-\frac{n}{2}+1}^{q_{1}=%
\frac{n}{2}-1}q_{1}j_{1\sigma _{q_{1}}}}.  \notag
\end{eqnarray}
Therefore, $\prod\limits_{q_{1}=-\frac{n}{2}+1}^{q_{1}=\frac{n}{2}%
-1}c_{q_{1},\sigma _{q_{1}}}^{\prime }=\prod\limits_{q_{2}=\frac{n}{2}%
-1}^{q_{2}=-\frac{n}{2}+1}c_{q_{2},\sigma _{q_{2}}}^{\prime }. $ Tracing
back, we find Eq. (\ref{cofa2}) is proved. Therefore, by the Laplace
expansion, we have proved that $\det (A_{\underline{j}})$ and $\det (A_{n-%
\underline{j}-1})$ are identical.

Till now, we find that on the left side of $R_{E},$ $\Psi
_{b}=\sum\limits_{j=0}^{n-1}\overline{\alpha }_{j}\overline{\Psi }_{j}$ with
$\overline{\alpha }_{j}=(-1)^{j}\alpha _{j}$ and $\alpha _{j}=\alpha
_{n-1-j}.$
\subsubsection{Quantization condition}\label{quancon}
We observe that all exponentially dissipating components vanish in the range of $x$ far away from $R_{E}$ and inside the
well of ${} V (x)$.
Only two free particle components with the form $e^{\pm ik_{0}}$ exist. The
phase factors of the free particle components are listed below:
\begin{equation}
\left(
\begin{array}{c}
\overline{\Psi }_{0} \\
\vdots \\
\overline{\Psi }_{j} \\
\vdots \\
\overline{\Psi }_{n-1}%
\end{array}%
\right) =\left(
\begin{array}{c|ccc}
\cdots & c_{j=0,q=+\frac{n}{2}} & c_{j=0,q=-\frac{n}{2}} & \cdots \\
\ddots & \vdots & \vdots & \ddots \\
\cdots & c_{j,q=+\frac{n}{2}} & c_{j,q=-\frac{n}{2}} & \cdots \\
\ddots & \vdots & \vdots & \ddots \\
\cdots & c_{j=n-1,q=+\frac{n}{2}} & c_{j=n-1,q=-\frac{n}{2}} & \cdots%
\end{array}%
\right) \left(
\begin{array}{c}
\vdots \\
\overline{\Psi }_{k_{q=+\frac{n}{2}}} \\
\overline{\Psi }_{k_{q=-\frac{n}{2}}} \\
\vdots%
\end{array}%
\right) .
\end{equation}%
Since $c_{j,q=\frac{n}{2}}=(e^{i\frac{\pi }{n}\ast q})^{-2b_{j}-\nu _{j}-%
\frac{1}{2}}=$ $e^{i[-\pi b_{j}-\pi (\frac{\nu _{j}}{2}+\frac{1}{4})]}$ and
$c_{j,q=-\frac{n}{2}}=(e^{i\frac{\pi }{n}\ast q})^{-2b_{j}-\nu _{j}-\frac{1}{%
2}}e^{i\pi (\nu _{j}+\frac{1}{2})}=e^{i[+\pi b_{j}+\pi (\frac{\nu _{j}}{2}+%
\frac{1}{4})]},$ using the last phases in the two formulas to define $%
\theta _{j}=\arg (c_{j,q=\frac{n}{2}})$ and $\theta _{j,q}=\arg (c_{j,q}),$
then we observe two relationships between these phases: $\theta _{j,q=\frac{n%
}{2}}=-\theta _{j,q=-\frac{n}{2}}$ and $\theta _{j}+\theta _{n-1-j}=\frac{%
\pi }{n}q(-j\frac{n}{n+1}+\frac{1}{n+1}-\frac{1}{2})+\frac{\pi }{n}q[-(n-1-j)%
\frac{n}{n+1}+\frac{1}{n+1}-\frac{1}{2}]=-\frac{n}{2}\pi +\frac{1}{2}\pi =-%
\frac{n}{2}\pi -\frac{1}{2}\pi +\pi $ since $c_{j,q=\frac{n}{2}}=(e^{i\frac{%
\pi }{n}\ast q})^{-j\frac{n}{n+1}+\frac{1}{n+1}-\frac{1}{2}}. $ The
bound state wave function inside the well becomes
\begin{eqnarray}
\Psi _{b} &=&\sum\limits_{j=0}^{n-1}\overline{\alpha }_{j}\overline{\Psi }%
_{j}=\sum\limits_{j=0}^{n-1}(-1)^{j}\alpha _{j}\overline{\Psi }_{j}  \notag
\\
&=&\sum\limits_{j=0}^{n-1}(-1)^{j}\alpha _{j}c_{j,q=+\frac{n}{2}}\overline{%
\Psi }_{k_{q=+\frac{n}{2}}}+(-1)^{j}\alpha _{j}c_{j,q=-\frac{n}{2}}\overline{%
\Psi }_{k_{q=-\frac{n}{2}}} \\
&\propto &\sum\limits_{j=0}^{n-1}(-1)^{j}\alpha _{j}\left[
\begin{array}{c}
c_{j,q=+\frac{n}{2}}\overline{\phi }_{0}^{\frac{1}{n+1}}\overline{\phi }%
_{0}^{\frac{-1}{2}}e^{i\int_{x}^{R_{E}}k_{0}dx^{\prime }}+ \\
c_{j,q=-\frac{n}{2}}\overline{\phi }_{0}^{\frac{1}{n+1}}\overline{\phi }%
_{0}^{\frac{-1}{2}}e^{i\int_{x}^{R_{E}}-k_{0}dx^{\prime }}%
\end{array}%
\right]  \notag \\
&=&\overline{\phi }_{0}^{\frac{1}{n+1}}\overline{\phi }_{0}^{\frac{-1}{2}%
}\sum\limits_{j=0}^{n-1}(-1)^{j}\alpha _{j}\left[ e^{i\theta
_{j}}e^{i\int_{x}^{R_{E}}k_{0}dx^{\prime }}+e^{-i\theta
_{j}}e^{i\int_{x}^{R_{E}}-k_{0}dx^{\prime }}\right]  \notag \\
&\propto &k_{0}^{\frac{1-n}{2}}\sum\limits_{j=0}^{n-1}(-1)^{j}\alpha _{j}%
\left[ \cos (\int_{x}^{R_{E}}k_{0}dx^{\prime }+\theta _{j})\right]  \notag \\
&=&k_{0}^{\frac{1-n}{2}}\frac{1}{2}\sum\limits_{j=0}^{n-1}\left\{
\begin{array}{c}
(-1)^{j}\alpha _{j}\left[ \cos (\int_{x}^{R_{E}}k_{0}dx^{\prime }+\theta
_{j})\right] + \\
(-1)^{n-1-j}\alpha _{n-1-j}\left[ \cos (\int_{x}^{R_{E}}k_{0}dx^{\prime
}+\theta _{n-1-j})\right]%
\end{array}%
\right\}  \notag \\
%&=&k_{0}^{\frac{1-n}{2}}\frac{1}{2}\sum\limits_{j=0}^{n-1}\left\{
%\begin{array}{c}
%(-1)^{j}\alpha _{j}\left[ \cos (\int_{x}^{R_{E}}k_{0}dx^{\prime }+\theta
%_{j})\right] + \\
%(-1)^{n-1-j}\alpha _{n-1-j}\left[ \cos (\int_{x}^{R_{E}}k_{0}dx^{\prime }+-%
%\frac{n}{2}\pi -\frac{1}{2}\pi +\pi -\theta _{j})\right]%
%\end{array}%
%\right\}  \notag \\
%&=&k_{0}^{\frac{1-n}{2}}\frac{1}{2}\sum\limits_{j=0}^{n-1}\left\{
%\begin{array}{c}
%(-1)^{j}\alpha _{j}\left[ \cos (\int_{x}^{R_{E}}k_{0}dx^{\prime }+\theta
%_{j})\right] + \\
%(-1)^{n-1-j+1}\alpha _{n-1-j}\left[ \cos (\int_{x}^{R_{E}}k_{0}dx^{\prime }-%
%\frac{n}{2}\pi -\frac{1}{2}\pi -\theta _{j})\right]%
%\end{array}%
%\right\}  \notag \\
&=&k_{0}^{\frac{1-n}{2}}\frac{1}{2}\sum\limits_{j=0}^{n-1}\left\{
\begin{array}{c}
(-1)^{j}\alpha _{j}\left[ \cos (\int_{x}^{R_{E}}k_{0}dx^{\prime }+\theta
_{j})\right] + \\
(-1)^{j}\alpha _{n-1-j}\left[ \cos (\int_{x}^{R_{E}}k_{0}dx^{\prime }-\frac{1%
}{2}\pi -\theta _{j})\right]%
\end{array}%
\right\}  \notag \\
&\overset{\alpha _{j}=\alpha _{n-1-j}}{=}&k_{0}^{\frac{1-n}{2}}\frac{1}{2}%
\sum\limits_{j=0}^{n-1}(-1)^{j}\alpha _{j}\left[ \cos
(\int_{x}^{R_{E}}k_{0}dx^{\prime }+\theta _{j})+\cos
(\int_{x}^{R_{E}}k_{0}dx^{\prime }-\frac{1}{2}\pi -\theta _{j})\right] .
\notag
\end{eqnarray}

In parallel, for the right side of $L_{E},$
\begin{eqnarray}
\Psi_b ^{L} &\propto &\overline{\phi ^{L}}_{0}^{\frac{1}{n+1}}\overline{\phi
^{L}}_{0}^{\frac{-1}{2}}\frac{1}{2}\sum\limits_{j=0}^{n-1}(-1)^{j}\alpha
_{j}^{L}\left[ \cos (\int_{L_{E}}^{x}k_{0}dx^{\prime }+\theta _{j})+\cos
(\int_{L_{E}}^{x}k_{0}dx^{\prime }-\frac{1}{2}\pi -\theta _{j})\right]
\notag \\
&\propto &k_{0}^{\frac{1-n}{2}}\frac{1}{2}\sum\limits_{j=0}^{n-1}(-1)^{j}%
\alpha _{j}^{L}\left[ \cos (\int_{L_{E}}^{x}k_{0}dx^{\prime }+\theta
_{j})+\cos (\int_{L_{E}}^{x}k_{0}dx^{\prime }-\frac{1}{2}\pi -\theta _{j})%
\right] .
\end{eqnarray}

The bound state is the linear combination of the two free particle
components in the region $L_{E}<x<R_{E}$ where the asymptotic formula is
valid. The linear combination coefficients are determined with the
provided $\Psi _{b}^{R}=\Psi_b $ and $\Psi _{b}^{L}$. Thus, everywhere in
this region satisfies both $\Psi _{b}^{R}$ and $\Psi _{b}^{L}.$ The only
possibility for the connection between $\Psi_b ^{L}$ and $\Psi_b ^{R}$ is
\begin{equation}
\int_{L_{E}}^{x}k_{0}dx^{\prime }+\theta _{j}
=-(\int_{x}^{R_{E}}k_{0}dx^{\prime }-\frac{1}{2}\pi -\theta _{j})+N\pi ,%
\text{i.e.,}
\int_{L_{E}}^{R_{E}}k_{0}dx^{\prime } =N\pi +\frac{1}{2}\pi ,  \notag
\end{equation}%
where $N=0,1,2,\cdots ,$ i.e. $N\in
%TCIMACRO{\U{2115} }%
%BeginExpansion
\mathbb{N}
%EndExpansion
_{0}. $

\subsection{$n=2\ast (2n_{c}+1)$ { {} with $n_c=0, 1, 2, \cdots$} }

The procedure parallels the previous section, and the
conclusion is identical. The only difference is the plus/minus symbol in front
of $\nabla ^{^{n}}\Psi $ {{} in the equation $ e^{-i\pi n/2}{{}\frac{d^n\Psi(x)}{d x^n} }
=[E-{{} V(x) }]\Psi(x) $}, which causes the exchange of the arguments of $k$
and $\kappa .$ On the right side of $R_{E},$ $\kappa =\kappa _{0},e^{\pm i%
\frac{2\pi }{n}\ast 1}\kappa _{0},e^{\pm i\frac{2\pi }{n}\ast 2}\kappa
_{0},\cdots ,\newline
e^{\pm i\frac{2\pi }{n}\ast n_{c}}\kappa _{0}$ correspond to the $n/2$
exponentially growing components; $\kappa =e^{\pm i\frac{2\pi }{n}\ast
(n_{c}+1)}\kappa _{0},\newline
e^{\pm i\frac{2\pi }{n}\ast (n_{c}+2)}\kappa _{0},\cdots ,e^{\pm i\frac{2\pi
}{n}\ast 2n_{c}}\kappa _{0},e^{i\frac{2\pi }{n}\ast n/2}\kappa _{0}$
indicate the $n/2$ evanescent components. On the left side of $R_{E},$ $%
k=e^{\pm i\frac{\pi }{n}\ast 1}k_{0},e^{\pm i\frac{\pi }{n}\ast
3}k_{0},\cdots ,e^{\pm i\frac{\pi }{n}\ast (2n_{c}-1)}k_{0}$ denote the $%
n/2-1$ exponentially growing components; $k=e^{\pm i\frac{\pi }{n}\ast
(2n_{c}+3)}k_{0},e^{\pm i\frac{\pi }{n}\ast (2n_{c}+5)}k_{0},\cdots ,e^{\pm i%
\frac{\pi }{n}\ast (n-1)}k_{0}$ correspond to the $n/2$ evanescent
components, and $e^{\pm i\frac{\pi }{n}\ast (2n_{c}+1)}k_{0}$ are free
particle components. In the matrix $A,$ $q$'s have these values: $%
-(n/2-1),-(n/2-2),\cdots ,0,\cdots ,(n/2-2),(n/2-1).$ The following
calculations are straightforward.

\section{Applications}

\subsection{Schr\"{o}dinger equation and Bogoliubov-de Gennes equation}

\bigskip $n=2$ case is the Schr\"{o}dinger equation ${} \frac{(-i\hbar
)^{2}}{2m}\frac{d^2\Psi }{dx^2}+V(x)\Psi =E\Psi $, and WKB gives exact energies of bound
states in the quantum harmonic oscillators.

$n=4$ case is the Bogoliubov de-Gennes (BdG) equation for the semimetals with parabolic dispersions, which describes
superconductors and reads as follows:
\begin{equation}
\left(
\begin{array}{cc}
H_{0}-E_{F} & {} \Delta(x)\\
{} \Delta(x)^\ast  & E_{F}-H_{0}%
\end{array}%
\right) \left(
\begin{array}{c}
\Psi \\
\Phi%
\end{array}%
\right) =E\left(
\begin{array}{c}
\Psi \\
\Phi%
\end{array}%
\right) ,
\end{equation}
with the original Hamiltonian $H_{0}=\left(
\begin{array}{cc}
0 & {} -\frac{\hbar ^{2}}{2m}\frac{d^2 }{dx^2} \\
{}  -\frac{\hbar ^{2}}{2m}\frac{d^2}{dx^2} & 0%
\end{array}%
\right) $, the superconducting order parameter ${} \Delta(x)$ , the Fermi energy $E_F\approx0$ , and the two-component wave functions $\Psi $ and $\Phi $. {{} Note that the superconducting order parameter is a complex function, and $\Delta(x)^\ast$ is the complex conjugate of $\Delta(x)$. Let $\Psi _{1}$ be the first component of $\Psi$. Considering that $\Delta(x)$ is a slowly-varying function, we can get $(\frac{\hbar ^{2}}{2m}%
)^{2}\partial_x ^{4}\Psi _{1}=(E^{2}-\Delta ^{2})\Psi _{1}$  by discarding the derivative terms of $\Delta(x)$. }We discuss the BdG
equation in details in another paper where
the notations have minor differences \cite{fan2022}.

\subsection{Multipartite Bogoliubov-de Gennes equations}
The BdG equation includes electrons and holes, i.e. two kinds of particles.
Inspired by this, we propose a series of multipartite BdG equations which
satisfy the formalism in this paper. Let $n=2^{n_{c}}$, and see below,
\begin{equation}
\left(
\begin{array}{cccccc}
-{} \frac{d^2 }{dx^2} & {} \Delta(x) & 0 & 0 & 0 & 0 \\
0 & {} \frac{d^2 }{dx^2}  & {} \Delta(x) & 0 & 0 & 0 \\
0 & 0 & \ddots & \ddots & 0 & 0 \\
0 & 0 & 0 & \ddots & {} \Delta(x) & 0 \\
0 & 0 & 0 & 0 & -\exp [i\pi (2^{n_{c}-1}-1)/2^{n_{c}-1}]{} \frac{d^2 }{dx^2}
& {} \Delta(x) \\
{} \Delta(x) & 0 & 0 & 0 & 0 & \exp [i\pi
(2^{n_{c}-1}-1)/2^{n_{c}-1}]{} \frac{d^2 }{dx^2}
\end{array}%
\right) \left(
\begin{array}{c}
\psi _{1} \\
\psi _{2} \\
\vdots \\
\psi _{n-1} \\
\psi _{n}%
\end{array}%
\right) =E\left(
\begin{array}{c}
\psi _{1} \\
\psi _{2} \\
\vdots \\
\psi _{n-1} \\
\psi _{n}%
\end{array}%
\right) ,
\end{equation}
where ${} \Delta(x) $ is the coupling strength between different parts and the
diagonal terms are the expansions of $E^{n}-\frac{d^{2n} }{dx^{2n}}$ When $%
{} \Delta (x)$ changes slowly, the equations can be reduced to our form. The
coupling strength between different parts can be non-identical, and we can
exchange the diagonal terms. The sole condition is keeping this system of
equations in a unidirectional ring. The multipartite BdG equation is
non-Hermitian, and the meaning is not decided yet.
% One shortcoming is that
%when ${} V =0,$ most parts do not have free particle solutions, which may
%hinder the application to physics.

\section{Remarks}
\begin{remark}
The hypothesis that all exponentially growing components are negligible is
appropriate because the well is assumed to be wide. Otherwise, the
asymptotic formulas are invalid. If these components are finite, the wave
function cannot be continuously patched within the well, or the wave function diverges
outside.
\end{remark}

\begin{remark}
The numerical wave functions are not available in our method except for the $%
n=2$ case, since we discard $O(1/z)$ terms in the Bessel functions.\cite%
{fan2022} The $O(1/z)$ terms multiplied by $e^{z}$ terms are much larger
than $e^{-z}$ terms, and in consequence, even a tiny deviation must destroy
numerical wave functions. Hence, the prerequisite of our approximation is a
hidden hypothesis that the bound states must exist. As for the $n=2$ case,
it exactly cancels all $O(1/z)$ terms coincidently. In the Schr\"{o}dinger
equation, the wave function is the linear combination of two Bessel
functions, for instance, $I_{+1/3}(z)-I_{-1/3}(z)$ in the classically
forbidden region. Since $\nu ^{2}=(\pm 1/3)^{2}=1/9,$ all the $e^{z}$ terms vanish according to the complete asymptotic expansion of $I_{\nu }(z):$%
\cite{watson1995}
\begin{equation}
I_{\nu }(z)\sim \frac{e^{z}}{(2\pi z)^{1/2}}\sum\limits_{m=0}^{\infty }\frac{%
(-1)^{m}(\nu ,m)}{(2z)^{m}}+\frac{e^{-z-(\nu +1/2)\pi i}}{(2\pi z)^{1/2}}%
\sum\limits_{m=0}^{\infty }\frac{(\nu ,m)}{(2z)^{m}},
\end{equation}%
with $(\nu ,m)=\frac{[4\nu ^{2}-1^{2}][4\nu ^{2}-3^{2}]\cdots \lbrack 4\nu
^{2}-(2m-1)^{2}]}{2^{2m}\cdot m!}$ and $(\nu ,0)=1$. To compare with the
numerical results, we obtain the eigenvalues and eigenfunctions in the $n=4$ case by
discretizing the original equation. The comparison is shown in  Appendix \ref{num}.
\end{remark}

\begin{remark}
When we use the asymptotic formulas of the Bessel functions, the phase may
be different for dissipating components as explained by the Stokes
phenomenon \cite{watson1995}. Nonetheless, the Stokes phenomenon
does not affect our proof here because these components vanish.
\end{remark}

\begin{remark}
\label{remarkdet}$\det (A)\neq 0$ is a conjecture. Whether the case $\det
(A)=0$ exists or not is not solved. If $\det (A)=0,$ the consequence is that
other bound states exist because the solution is not unique.
\end{remark}

\begin{remark}
In general, for odd $n,$ these bound states do not exist. The reason is that
only one free particle component exists in each region. If we let all
exponential growing components and a free particle component in the
classically forbidden region be zero, we will get $n$ equations for $n$
variables. In general, these equations do not have a nontrivial solution
unless the determinant of the coefficient matrix is zero. Whether such
special cases exist is not solved. Or, we let the free particle component
exist in the classically forbidden region, then the spectrum is continuous,
and no bound states exist.
\end{remark}

%\begin{remark}\label{remarkdouble}
%As the conclusion is given, we provide an alternative perspective. We consider $e^{-i\pi n/2}\nabla_x ^{^{n}}$ and $E-{} V $ as two operators, and take an approximate $n/2$th root of both operators: $(e^{-i\pi n/2}\nabla_x ^{^{n}})^{2/n}\sim -\nabla_x ^2$ and $(E-{} V)^{2/n} \sim \text{sign}(E-{} V)|E-{} V|^{2/n}$. Then all cases are reduced to the Schr\"{o}dinger-like equation which can be solved by the traditional WKB method. According to this reduced form, the rules of the double-well problem in the Schr\"{o}dinger equation perhaps apply to the $n>2$ cases. In Appendix \ref{dwp}, we give numerical examples for this discussion.
%\end{remark}
\begin{remark}\label{remarkdouble}
The double-well problem for the $n$th order Schr\"{o}dinger equation is not solved directly. In Appendix \ref{dwp}, we give numerical discussions for the case of $n=4$.
\end{remark}

\begin{remark}
As in the Schr\"{o}dinger equation, e.g., Reference \cite{simon1983semiclassical}, the further study of the eigenvalues and eigenfunctions in these equations with higher $n$ can be topics in functional analysis. One can also try to add corrections to the current form as in the Schr\"{o}dinger equation. \cite{bender1999}
\end{remark}

\begin{remark}
It is intriguing that the phase factor echoes in this equation after
building several physical milestones such as the Ahanov-Bohm effect, Josephson
effect, Berry phase, etc. Furthermore, symmetry and quantization also play a central role in this paper.
\end{remark}
%\begin{acknowledgement}
%\end{acknowledgement}
\section*{Data availability}
 All data for numerical simulations are available from the author upon request.
\section*{Acknowledgments}

We thank YMSC and BIMSA for open lectures, thank Prof. X. Dai for stimulating discussions, thank Dr. Y. S. An, Prof. H. Q. Lin, and Prof. Y. Y. Li for some communications, and thank Prof. S. N. Tan for comments. The author also benefits from Prof. C. P. Sun's inspiring lectures. This work is partly supported by the Hong
Kong Research Grants Council (Project No. GRF16300918 and No. 16309020).
%\begin{thebibliography}{9}
%\bibitem{}
%*
%\end{thebibliography}
%\bibliographystyle{plain}
%\bibliographystyle{aipauth4-1}
%\bibliographystyle{unsrt}%{plain}
%\bibliographystyle{prsty}plain
%\bibliographystyle{amsplain}
\bibliographystyle{unsrt}
\bibliography{nusup}
\appendix
\setcounter{figure}{0}
\setcounter{table}{0}
\section{Notations}
\label{symbols} %~\\
\begin{equation}
\begin{array}{ccc}
\text{symbol} & \text{meaning} & \text{first appear around} \\
A_{\underline{j}} & \text{matrix obtained by replacing }\underline{j}\text{%
th column of }A\text{ with }b & \text{Eq. (40)} \\
A_{n-\underline{j}-1} & \text{matrix obtained by replacing }n-\underline{j}-1%
\text{th column of }A\text{ with }b & \text{Eq. (40)} \\
\alpha _{j},\overline{\alpha }_{j} & \text{coefficients in }\Psi _{b}=\sum
\alpha _{j}\Psi _{j},{\Psi }_{b}=\sum \overline{\alpha }_{j}\overline{\Psi }%
_{j} & \text{Eq. (33)} \\
c_{j,q},\overline{c}_{j,\overline{q}} & \text{phase factor with the
parameters }j,q & \text{Eq. (28), Eq. (32)} \\
j & \text{index for the different solutions around the turning points} &
\text{Eq. (2.5)} \\
\underline{j} & \text{column index of the matrix }A & \text{Eq. (2.40)} \\
k_{0} & \text{wave vector in the classically allowed region} & \text{Eq. (2.31)} \\
k & \text{wave vector or index} & \text{Eq. (2.32) or Eq. (2.8)}
\\
\kappa _{0} & \text{wave vector in the classically forbidden region} &
\text{Eq. (2.2)} \\
\kappa & \text{wave vector} & \text{Eq. (2.2)} \\
\phi _{0},\overline{\phi }_{0} & \text{phase variable corresponding to }%
\kappa _{0},k_{0} & \text{Eq.(2.8), Eq.(2.31)} \\
\phi _{q},\overline{\phi }_{q} & \text{phase variable corresponding to }%
\kappa ,k & \text{Eq.(2.26), Eq. (2.32)} \\
\Psi _{b} & \text{bound state wave function} & \text{Eq. (2.32)} \\
\Psi _{j},\overline{\Psi }_{j} & \text{wave functions obtained by the
Frobenius method} & \text{Eq. (2.5), Eq. (2.31)} \\
\Psi _{\kappa } & \text{asymptotic component in }\Psi _{j} & \text{Eq. (2.28)} \\
\overline{\Psi }_{k} & \text{asymptotic component in }\overline{\Psi }_{j} &
\text{Eq. (2.32)} \\
q,\overline{q}\text{ or just }q & \text{the parameter in the wave vector }%
\kappa ,k & \text{Eq. (2.23)\&Eq. (2.32)} \\
\overline{q} & \text{the parameter in the wave vector }k,\text{also used as
the row index} & \text{Eq. (2.32), Eq. (2.49)} \\
x,x^{\prime } & \text{normal Cartesian coordinate and the substitutions} &
\text{Eq. (2.1), Eq. (2.3), Eq. (2.29), Eq. (%
2.33)} \\
&  &  \\
&  &  \\
&  &
\end{array}%
\end{equation}

\section{ Proof of the linear approximation Eq. (2.2)\label{firsta}}
{{}
The first-order derivative terms are $n\kappa ^{n-1}\frac{d^{}f }{dx^{}}+\frac{n(n-1)}{%
2}f\kappa ^{n-2}\frac{d^{}\kappa }{dx^{}} ,$ and zeroth derivative terms cancel each
other. Now we consider other terms. In the first time the operator $\frac{d^{} }{dx^{}}$
acts on the hypothesized wave function, we get terms proportional to $\kappa
$ and $\frac{d^{}f }{dx^{}};$ then $\frac{d^{} }{dx^{}} $ acts again on the previous result, we get
terms $\kappa ^{(1)},$ $\frac{d^{2} f}{dx^{2}}$, $\kappa ^{2}$ and $\kappa \frac{d^{} f}{dx^{}};$
and so on and so forth. Based on this principle, we can enumerate all
possible terms up to a coefficient after the action of $\frac{d^{n} }{dx^{n}}$. We
express these terms as $f^{(d_{1})}\kappa ^{d_{2}}\frac{d^{d_3} \kappa}{dx^{d_3}} $
with $\left\{
\begin{array}{ll}
d_{1}+d_{2}+d_{3}=n, & \hbox{$d_{3}=0;$} \\
d_{1}+d_{2}+d_{3}+1=n, & \hbox{$d_{3}\neq 0;$}%
\end{array}
\right.$ $\frac{d^{0} \kappa}{dx^{0}}  =1;$ and $d_{1}+d_{3}>1.$
%Because we always
%assume that the potential varies slowly, we have two features: $\frac{%
%f^{(d_{1})}}{\frac{d^{} f}{dx^{}}}\ll 1$ with $d_{1}>1;$ and $\frac{\frac{d^{d_3} \kappa }{dx^{d_3}} } {\frac{d^{} \kappa}{dx^{}} }\ll 1$ with $d_{3}>1. $ For the
%asymptotic behavior, $\kappa \gg 1. $
In the following, we give the conditions that we can do the linear approximation in Eq. (2.2).
If $d_{1}=0,$ we get
\begin{equation}
\frac{f\kappa ^{d_{2}}\frac{d^{d_3} \kappa }{dx^{d_3}} }{f\kappa ^{n-2}\frac{d^{}\kappa }{dx^{}} }%
\propto \frac{\kappa ^{d_{2}}\frac{d^{d_3} \kappa }{dx^{d_3}}}{\kappa ^{n-2}\frac{d^{}\kappa }{dx^{}} }\ll 1\text{ with }d_{2}<n-2\text{ and }d_{3}\geq 2;
\end{equation}%
if $d_{1}=1$, we have
\begin{equation}
\frac{f^{(d_{1})}\kappa ^{d_{2}}\frac{d^{d_3} \kappa }{dx^{d_3}} }{\kappa ^{n-1}\frac{d^{}f }{dx^{}}
}\propto \frac{\kappa ^{d_{2}}\frac{d^{d_3} \kappa }{dx^{d_3}} }{\kappa ^{n-1}}\ll 1%
\text{ with }d_{2}<n-1\text{ and }d_{3}\geq 1;
\end{equation}%
if $d_{3}=0,$ we obtain
\begin{equation}
\frac{f^{(d_{1})}\kappa ^{d_{2}}}{\kappa ^{n-1}\frac{d^{}f }{dx^{}}}\ll 1\text{ with }%
d_{1}\geq 2\text{ and }d_{2}<n-1;
\end{equation}%
if $d_{3}=1$, we have
\begin{equation}
\frac{f^{(d_{1})}\kappa ^{d_{2}}\frac{d^{d_3} \kappa }{dx^{d_3}} }{f\kappa ^{n-2}\frac{d^{}\kappa }{dx^{}} }\propto \frac{f^{(d_{1})}\kappa ^{d_{2}}}{f\kappa ^{n-2}}\ll 1\text{
with }d_{1}\geq 1\text{ and }d_{2}<n-2;
\end{equation}%
otherwise, the result is
\begin{equation}
\frac{f^{(d_{1})}\kappa ^{d_{2}}\frac{d^{d_3} \kappa }{dx^{d_3}} }{f\kappa ^{n-2}\frac{d^{}\kappa }{dx^{}} }\ll 1\text{ with }d_{1}>1,\text{ }d_{2}<n-2\text{ and }d_{3}>1.
\end{equation}

Hence, {{} all the terms with order greater than one can be neglected when the above conditions are satisfied} and then Eq. (\ref{kappa0}) is proper for the asymptotic formula.}

\section{Proof of Eq. (2.23)}
\label{repre} Eq. (2.23): $\sum\limits_{q=\pm 1,\text{ }\pm
3,\cdots ,\text{ }\pm (\frac{n}{2}-1)}\frac{1}{n/2}(e^{i\frac{\pi }{n}\ast
q})^{m}=\left\{
\begin{array}{ll}
0, & \hbox{\text{ mod}$(m,n)\neq 0$;} \\
+1, & \hbox{\text{ mod}$(m/n,2)=0$;} \\
-1, & \hbox{\text{ mod}$(m/n,2)=1$.}%
\end{array}%
\right. $

This equation can be easily proved by the representation theory with a few
modifications. It is clear that
\begin{eqnarray}
\sum\limits_{q=\pm 1,\text{ }\pm 3,\cdots ,\text{ }\pm (\frac{n}{2}-1)}(e^{i%
\frac{2\pi }{n}\ast q})^{m/2} &=&\sum\limits_{q=\pm 1,\text{ }\pm 3,\cdots ,%
\text{ }\pm (\frac{n}{2}-1)}e^{i\left[ \frac{2\pi }{n}(m/2)\right] \ast q} \\
&=&e^{i\left[ \frac{2\pi }{n/2}(m/4)\right] \ast \lbrack -(\frac{n}{2}%
-1)]}\sum\limits_{q=\pm 1,\text{ }\pm 3,\cdots ,\text{ }\pm (\frac{n}{2}%
-1)}e^{i\left[ \frac{2\pi }{n/2}(m/4)\right] \ast \lbrack q+(\frac{n}{2}-1)]}
\notag \\
&=&e^{i\left[ \frac{2\pi }{n/2}(m/4)\right] \ast \lbrack -(\frac{n}{2}%
-1)]}\sum\limits_{q^{\prime }=0,\text{ }2,\cdots ,\text{ }n-2}e^{i\left[
\frac{2\pi }{n/2}(m/2)\right] \ast q^{\prime }/2}  \notag \\
&=&e^{i\left[ \frac{2\pi }{n/2}(m/4)\right] \ast \lbrack -(\frac{n}{2}%
-1)]}\sum\limits_{q^{\prime \prime }=0,\text{ }1,\cdots ,\text{ }\frac{n-2}{2%
}}e^{i\left[ \frac{2\pi }{n/2}(m/2)\right] \ast q^{\prime \prime }}.
\notag
\end{eqnarray}%
Treat $e^{i\left[ \frac{2\pi }{n/2}(m/2)\right] \ast q^{\prime \prime }}$
terms as the multiplications of different irreducible representations in the
equivalence classes of the cyclic group $Z_{n/2}$, then according to the
orthogonality theorem,\cite{zee2016group}
\begin{equation}
\sum\limits_{q^{\prime \prime }=0,\text{ }1,\cdots ,\text{ }\frac{n-2}{2}%
}e^{i\left[ \frac{2\pi }{n/2}(m/2)\right] \ast q^{\prime \prime
}}=\sum\limits_{q=\pm 1,\text{ }\pm 3,\cdots ,\text{ }\pm (\frac{n}{2}%
-1)}e^{i\left[ \frac{2\pi }{n}(m/2)\right] \ast q}=0,
\end{equation}%
when $-2\pi <\frac{2\pi }{n/2}(m/2)<2\pi ,$ i.e. $-n<m<n$. This conclusion
can be extended to the broader range mod$(m,n)\neq 0.$ First we need to deal
with mod$(m,n)=0$ separately. When mod$(m/n,2)=0,$
\begin{equation}
\sum\limits_{q=\pm 1,\text{ }\pm 3,\cdots ,\text{ }\pm (\frac{n}{2}-1)}(e^{i%
\frac{\pi }{n}\ast q})^{m}=n/2;
\end{equation}%
when mod$(m/n,2)=1,$
\begin{equation}
\sum\limits_{q=\pm 1,\text{ }\pm 3,\cdots ,\text{ }\pm (\frac{n}{2}-1)}(e^{i%
\frac{\pi }{n}\ast q})^{m}=-n/2.
\end{equation}%
Otherwise, for mod$(m,n)\neq 0,$ we can always separate a multiple of $\ n$
from $m$ with the remainder $m^{\prime }$ so that $-n<m^{\prime }<n,$ on
this account, we have
\begin{equation}
\sum\limits_{q=\pm 1,\text{ }\pm 3,\cdots ,\text{ }\pm (\frac{n}{2}-1)}(e^{i%
\frac{\pi }{n}\ast q})^{m}=\sum\limits_{q=\pm 1,\text{ }\pm 3,\cdots ,\text{
}\pm (\frac{n}{2}-1)}\pm 1(e^{i\frac{\pi }{n}\ast q})^{m^{\prime }}=0.
\end{equation}

Straightforwardly, we get the paralleled function substituted in Eq. (2.32).

\section{Numerical results for $n=4$}\label{num}
The following table shows the eigenvalue $E$ in the numerical results and
our method with the harmonic oscillator-type potential well ${} V
=(1/2)\alpha x^{2}$. By discretizing the original equation, the eigenvalues
are obtained by diagonalizing the corresponding matrix. The cut-offs of
numerical calculations are $x=\pm 40$, and the dimensions of the
discretization are $6000$.

From our formula, the eigenvalue is
\begin{equation}
E=[(N+1/2)\frac{3\sqrt{\pi }(\alpha /2)^{1/2}\Gamma (3/4)}{\Gamma (1/4)}]^{4/3}.
\end{equation}%
From \ref{tab:table1}, we see that as the number of bound states
increases, our approximate results approach the numerical eigenvalues
quickly. This performance is not surprising considering that the WKB result is
exact for the quantum harmonic oscillator. The ratios between the exact results and the WKB results are identical with varying $\alpha $ according to scale transformations.

\begin{figure}[bh]
%\captionstyle{centerlast}
%\includegraphics[clip,scale=0.5, angle=0]{wzj3}
\includegraphics[clip,scale=0.500, angle=0]{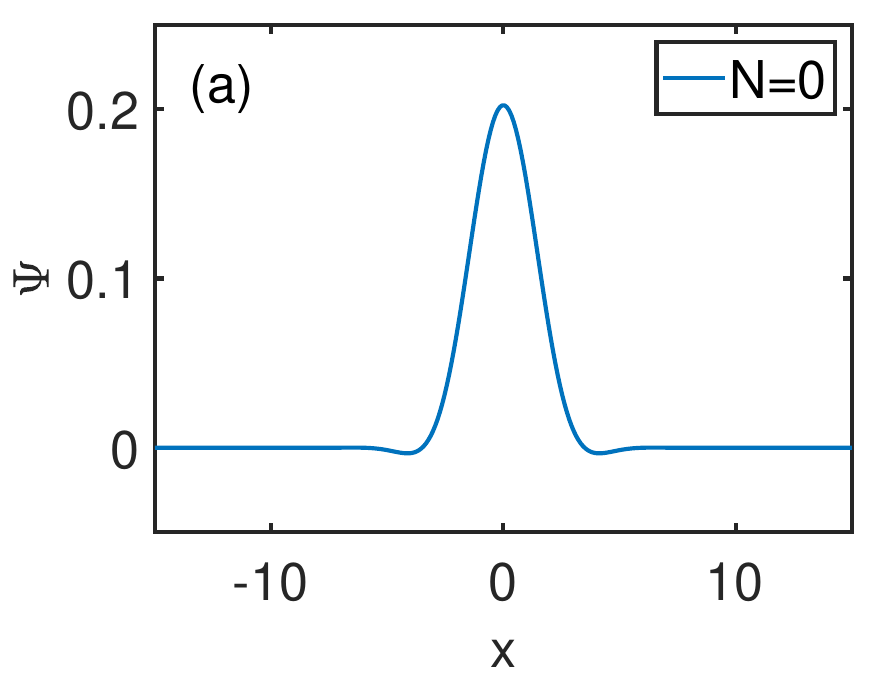}
\includegraphics[clip,scale=0.500, angle=0]{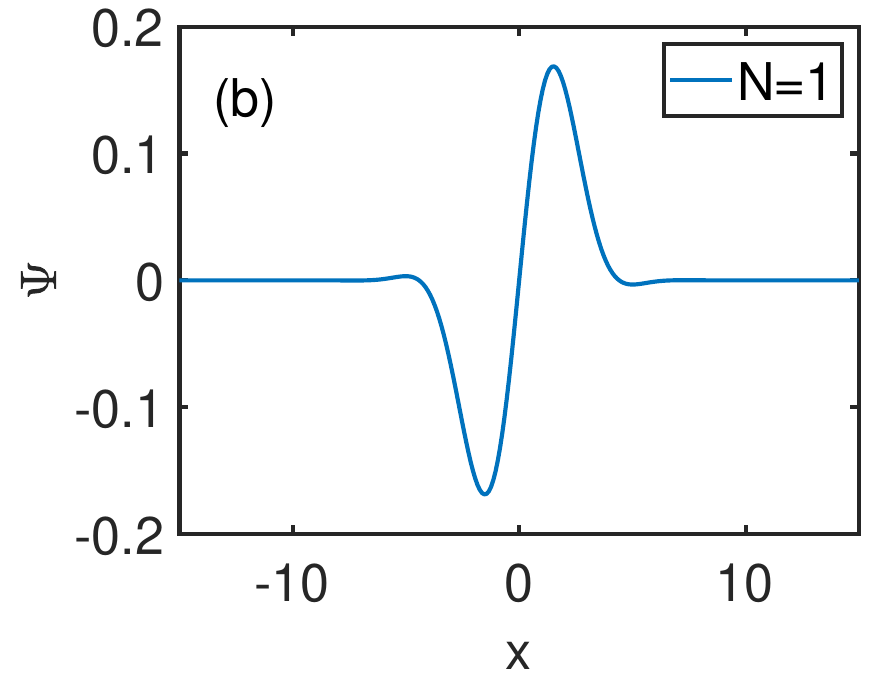}
\includegraphics[clip,scale=0.500, angle=0]{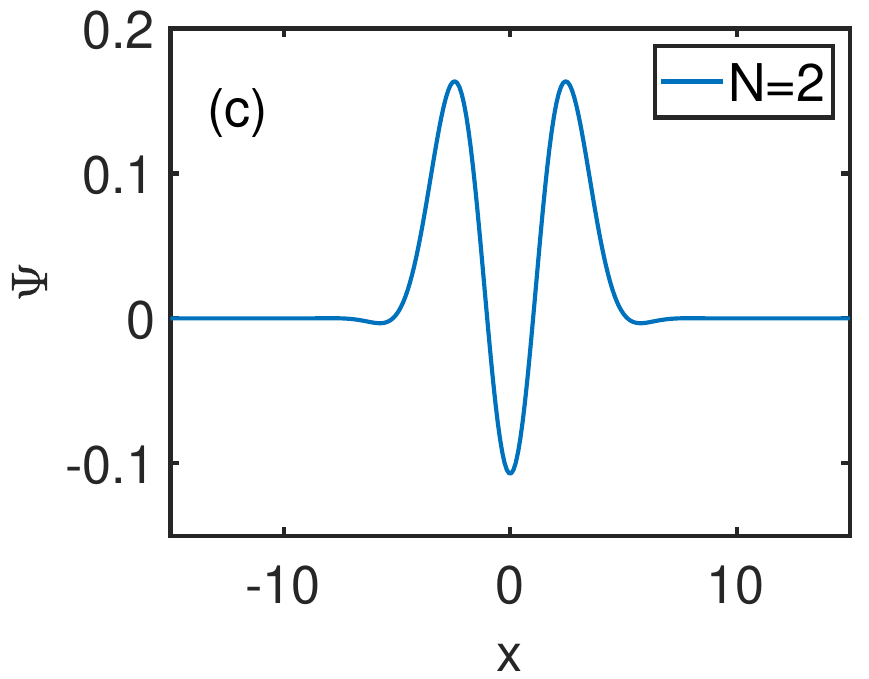}
\includegraphics[clip,scale=0.50, angle=0]{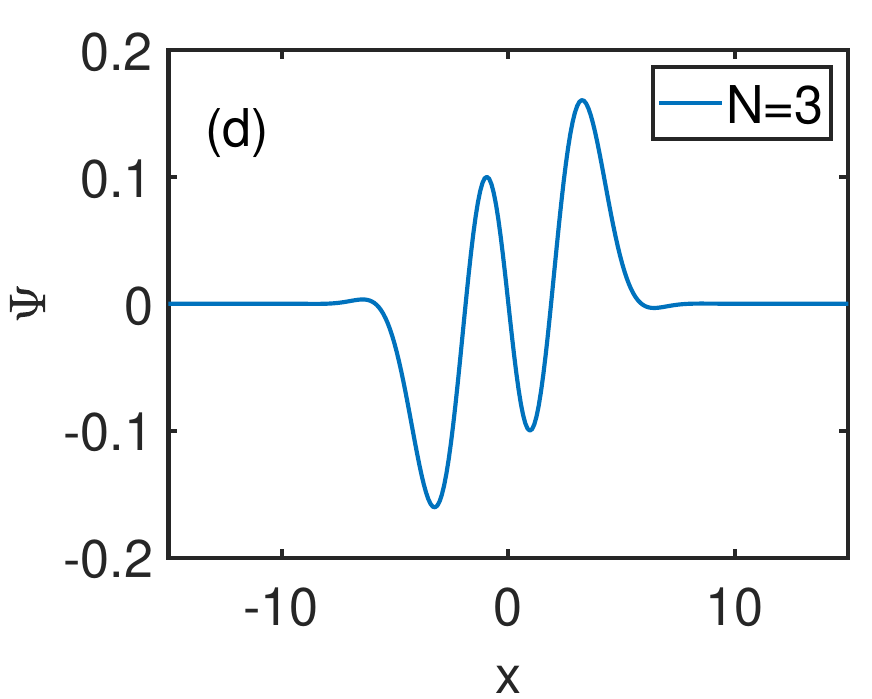}
\caption{The first four bound states for $n=4$. Note that there are tails or small oscillations when the wave functions are damped to zero, which are the high-order effect and absent in the Schr\"{o}dinger equation as expected from our asymptotic analysis.}
\label{BS4}
\end{figure}

The first four bound states wave functions are shown in \ref{BS4}. Akin to the quantum harmonic oscillator, the $N$th bound state has $N$ node(s) or zero(s) inside the well, and the parity changes as $N$ adds or decreases one. The tails at two sides are caused by the oscillating damped terms, and the nodes caused by these are not counted. We observe that the tails, which are absent in the Schr\"{o}dinger equation, are the high-order effect.
\begin{subappendices}
\begin{table}[bh]
\begin{center}
\caption{Numerical and WKB results for $n=4$.}\label{tab:table1}
\begin{tabular}{cccccccccccc}
No. of states &0 &1&2&3&4&5&6 \\
NUM $\alpha=1 $&0.6680 &2.3936 &4.6967 &7.3356 &10.2440 &13.3789 & 16.7113 \\
WKB $\alpha=1 $	&0.5463 &2.3636 & 4.6705 &7.3148 &10.2265 &13.3638 &16.6979 \\
No. of states &7&8&9&10&11&12&13 \\
NUM $\alpha=1 $&20.2201&23.8890 &27.7052 & 31.6580&35.7386 &39.9394 &44.2538 \\
WKB $\alpha=1 $	&20.2082 &23.8783 &27.6956 &31.6493&35.7308 &39.9324 &44.2477\\
No. of states &14&15&16&17&18&19&20 \\
NUM $\alpha=1 $&48.6763 &53.2017 &57.8256 &62.5439 &67.3530 &72.2495 &77.2306 &\\
WKB $\alpha=1 $	&48.6709 &53.1970 &57.8216 &62.5407 &67.3505 &72.2478 &77.2296 &\\
\end{tabular}
\end{center}
\end{table}

%\begin{figure}[tbph]
%\centering
%
%% Requires \usepackage{graphicx}
%\includegraphics[width=9cm]{ratio.pdf}
%\caption{Ratios between the numerical results and our results versus the number of states. Different $\alpha$ cases coincide. The closer the ratio is from the unity, the better the approximation is.}\label{alpha}
%\end{figure}

\end{subappendices}
\section{Double-well problem}\label{dwp}
\begin{figure}[!h]
%\captionstyle{centerlast}
%\includegraphics[clip,scale=0.5, angle=0]{wzj3}
\includegraphics[clip,scale=0.500, angle=0]{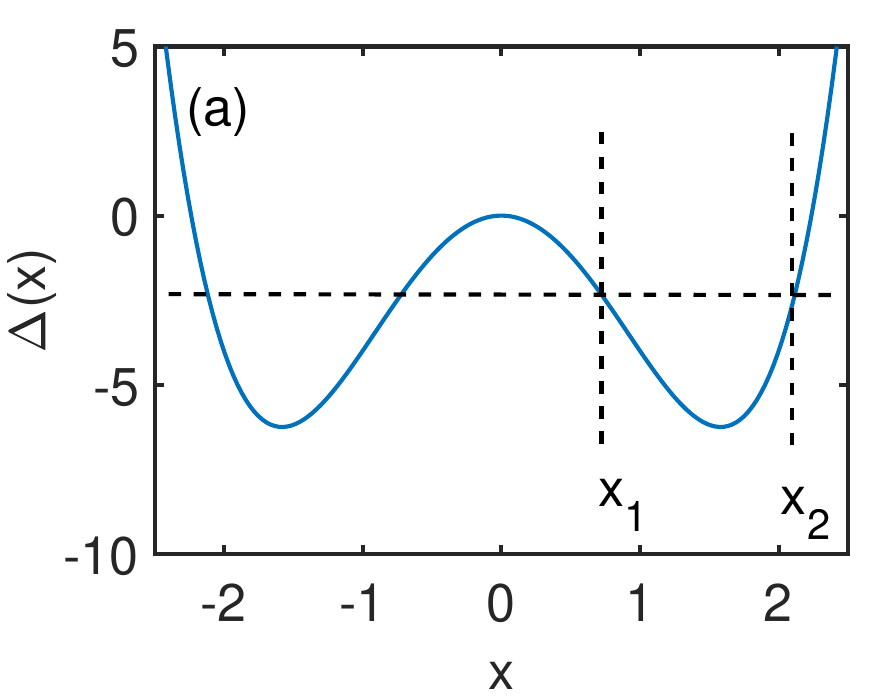}
\includegraphics[clip,scale=0.500, angle=0]{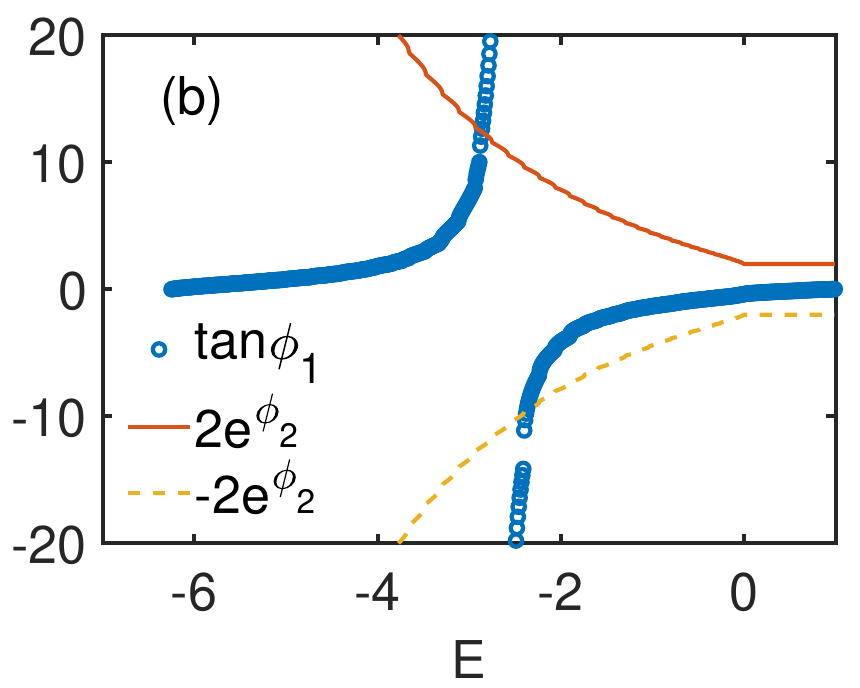}
\includegraphics[clip,scale=0.500, angle=0]{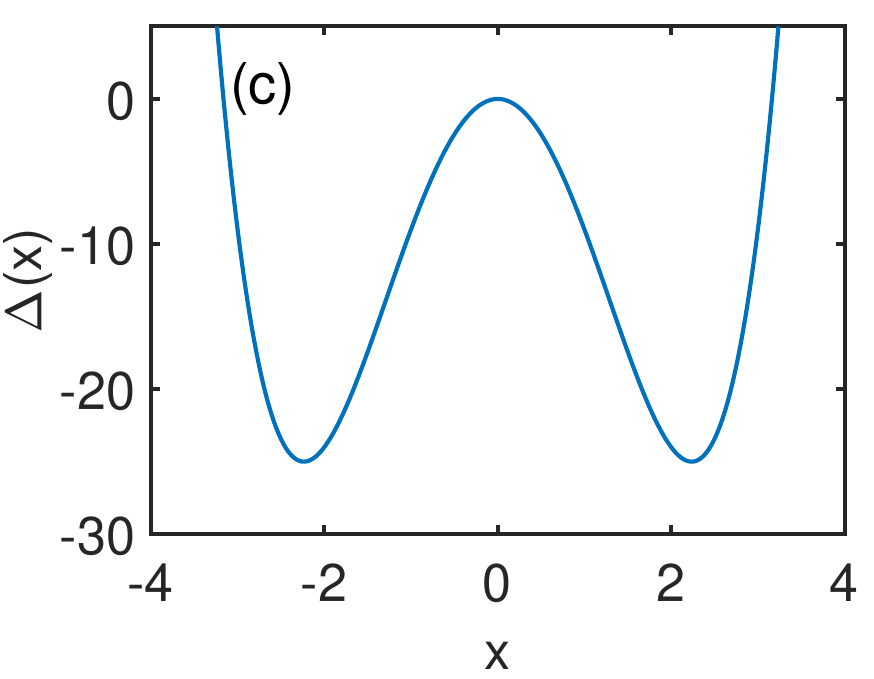}
\includegraphics[clip,scale=0.50, angle=0]{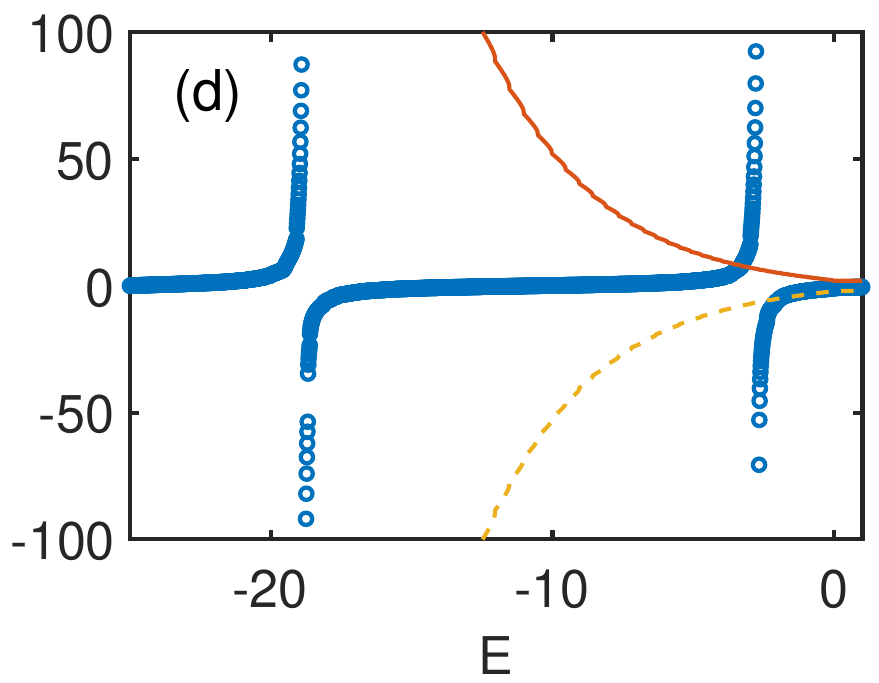}
\caption{(a) The double-well potential for ${{} V(x) }=x^4-5x^2$; (b) the intersection points between $\tan \phi_1$ and $\pm2e^{\phi_2}$ are the bound states in (a). $x_1$ and $x_2$ are the intersection points of $E$ and ${{} V(x) }$ on the right-side well. (c)  The double-well potential for ${{} V(x) }=x^4-10x^2$. (d) The intersection points between $\tan \phi_1$ and $\pm2e^{\phi_2}$ are the bound states in (c). n=4 for all plots.}
\label{double}
\end{figure}

The double-well problem is not directly solved. However, based on the discussion in this paper, we may borrow the formulas in the Schr\"{o}dinger equation \cite{griffiths2018} to solve the double-well problem in $n>2$ cases. Numerically, we find that the WKB formulas in the Schr\"{o}dinger equation perform well in solving the eigenvalues of the bound states in higher-order differential equations. As shown in \ref{double}, we discuss the double-well problem for $n=4$. Define $\phi_1=\int_{x_1}^{x_2}k_0dx$ and $\phi_2=\int_{-x_1}^{x_1}\kappa_0dx$ where $x_1$ and $x_2$ are the intersection points of $E$ and ${{} V(x) }$ on the right-side well as explained in the figure. The bound states are determined by $\tan \phi_1=\pm2e^{\phi_2}$, which can be solved numerically. $\pm$ in the Schr\"{o}dinger equation corresponds to the even/odd parity of the wave function. However, this correspondence becomes ambiguous for $n=4$ according to our numerical data which are not shown. Despite the ambiguity of the parity of wave functions, the eigenvalues are well captured. Collecting the intersections points in \ref{double}, we list the numerical results and the WKB results in \ref{tab:double}. As seen the error is small, and the WKB approximation works well.

\begin{subappendices}
\begin{table}[!h]
\begin{center}
\caption{Numerical and WKB results for the double-well problem with $n=4$.}\label{tab:double}
\begin{tabular}{ccc|cccccccc}
\multicolumn{3}{c|}{${{} V(x) }=x^4-5x^2$} &\multicolumn{4}{|c}{${{} V(x) }=x^4-10x^2$} \\
No. of states &0 &1&0&1&2&3 \\
parity &even &odd  &odd &even &even & odd \\
NUM &-2.4846 &-1.9661  &-17.7793 &-17.6288 &-3.9839 & -1.9689 \\
WKB &-2.8793 &-2.4093 &-18.9096 &-18.7496 &-3.3296 &-2.0496 \\
%NUM/WKB &0.8629& 0.8160&0.9402& 0.9402& 1.1965& 0.9606 \\
\end{tabular}
\end{center}
\end{table}
\end{subappendices}

\end{document}